\documentclass[reqno]{amsart}

\usepackage{amssymb}

\newtheorem{theorem}{Theorem}[section]
\newtheorem{proposition}[theorem]{Proposition}
\newtheorem{lemma}[theorem]{Lemma}

\newtheorem{definition}[theorem]{Definition}
\newtheorem{remark}[theorem]{Remark}

\numberwithin{equation}{section}

\newenvironment{acknowledgement}{\emph{Acknowledgement.}}

\DeclareMathOperator{\supp}{supp}

\DeclareMathOperator{\dist}{dist}

\newcommand{\pr}{\prime}

\renewcommand\L{\mathrm{L}}
\newcommand\R{\mathbb R}
\newcommand\N{\mathbb N}

\newcommand\Z{\mathbb Z}

\newcommand\G{\mathbb{G}}

\newcommand\di{\mathrm d}

\newcommand\X{\mathbf{X}}
\newcommand\Y{\mathbf{Y}}
\newcommand\beps{\boldsymbol{\varepsilon}}

\newcommand\J{\mathbb J}

\renewcommand\S{\mathcal{S}}
\newcommand{\cX}{\mathcal{X}}
\newcommand{\cP}{\mathcal{P}}
\newcommand{\cQ}{\mathcal{Q}}
\newcommand{\cC}{\mathcal{C}}
\newcommand{\cJ}{\mathcal{J}}
\newcommand{\cD}{\mathcal{D}}
\newcommand{\cR}{\mathcal{R}}
\newcommand{\cG}{\mathcal{G}}

\newcommand\e{\mathrm{e}}

\newcommand\eps{\varepsilon}
\newcommand{\vrho}{\varrho}

\newcommand\vphi{\varphi}

\newcommand{\la}{\langle}
\newcommand{\ra}{\rangle}

\renewcommand\P{\mathbb P}

\newcommand{\abs}[1]{\left\lvert #1 \right\rvert}
\newcommand{\norm}[1]{\left\lVert #1 \right\rVert}
\newcommand{\scal}[1]{\la #1 \ra}

\newcommand\beq{\begin{equation}}
\newcommand\eeq{\end{equation}}

\begin{document}


\title[Localization for   Poisson random Schr\"odinger operators]
{Localization for  Schr\"odinger operators with  Poisson random potential}

 \author{Fran\c cois Germinet}
  \address{ Laboratoire AGM, CNRS UMR 8088, D\'epartement de Math\'ematiques,
   Universit\'e de Cergy-Pontoise,
  Site de Saint-Martin,
  2 avenue Adolphe Chauvin,
 95302 Cergy-Pontoise cedex, France}
  \email{germinet@math.u-cergy.fr}

\author{Peter D. Hislop}
\address{ Department of Mathematics, University of Kentucky, 
Lexington, KY 40506-0027, USA}
 \email{hislop@ms.uky.edu}

\author{Abel Klein}
\address{University of California, Irvine,
Department of Mathematics,
Irvine, CA 92697-3875,  USA}
 \email{aklein@uci.edu}

\thanks{P.H.  was  supported in part  by NSF Grant
DMS-0503784.}
\thanks{A.K was  supported in part by NSF Grant DMS-0457474.}

\begin{abstract}
We prove exponential and dynamical
localization for the  Schr\"oding\-er operator with a
nonnegative Poisson random potential at the bottom of the
spectrum in any dimension.
We also conclude that the eigenvalues in that spectral region of localization
have finite multiplicity.
We prove similar localization results
in a prescribed energy
interval at the bottom of the
spectrum provided the density of the Poisson process is large enough.
\end{abstract}

\maketitle

\tableofcontents

\newpage

\section{Introduction and main results} 

\subsection{Background and motivation}

Consider an electron moving in an amorphous medium with randomly placed identical impurities, each impurity creating a local  potential.   For a fixed configuration of the impurities, described by the countable set $X\subset\R^{d}$  giving their locations,
this motion  is described by
the Schr\"odinger equation  $-i \partial_t \psi_{t} =
H_X \psi_{t}$ with the  Hamiltonian
\begin{gather}\label{PoissonH}
H_X :=  -\Delta + V_{X} \quad \text{on} \quad \L^{2}(\R^{d}) , \\
\intertext{where the potential is given by}
  V_{X}(x):=  \sum_{\zeta \in X} u(x - \zeta), \label{PoissonV}
\end{gather}
with  $u(x - \zeta)$ being the single-site potential created by the impurity placed at $\zeta$.  Since the impurities are randomly distributed, the configuration $X$ is a random countable subset of  $\R^{d}$, and hence it is modeled by a point process on $\R^{d}$.
Physical considerations usually dictate
that the process is homogeneous and ergodic with respect to the
translations by $\R^d$, cf.\ the discussions in \cite{LGP,PF}.
The canonical point process with the desired properties is the
homogeneous Poisson point process on $\R^{d}$.

The \emph{Poisson Hamiltonian} is  the random Schr\"odinger operator $H_{\X}$ 
in \eqref{PoissonH} with $\X$   a Poisson process on $\R^d$ with density  $\varrho >0$. The potential
 $ V_{\X}$ is then  a   \emph{Poisson random potential}.
 Poisson Hamiltonians may be  the most natural random Schr\"odinger
operators in the continuum as the distribution of impurities
in  various samples of material is naturally modeled by a Poisson
process.
A mathematical proof of the existence of localization in two or more dimensions has been a long-standing open problem (cf. the survey \cite{LMW}).  The Poisson Hamiltonian has been long known to have Lifshitz tails \cite{DV,CL,PF,Klop97,Sz,KP,St99}, a strong indication of localization at the bottom of the spectrum.  Up to now  localization had  been shown only in one dimension \cite{Stolz}, where it holds at all energies, as expected. 

In this article we prove  localization  for nonnegative Poisson Hamiltonians at the bottom of the spectrum in arbitrary dimension.  We obtain both exponential (or Anderson) localization and dynamical localization, as well as finite multiplicity of eigenvalues.  In a companion paper  \cite{GHK2} we modify our methods to obtain localization at low energies for  Poisson Hamiltonians with attractive (nonpositive) single-site potentials.

In the multi-dimensional continuum case localization has been shown in the case where   
 the randomness is given by  random variables with bounded densities.
There is a wealth of results concerning localization 
for Anderson-type Hamiltonians,   which
are $\Z^{d}$-ergodic random  Schr\"odinger operators as in \eqref{PoissonH}
but for which the location of the impurities is fixed
at the vertices of the lattice $\Z^{d}$ (i.e.,   $X\equiv\Z^{d}$), and the single-site potentials are multiplied by random variables with bounded densities, e.g.,    \cite{HM,CH,Klop95,KSS,Klop02,GKgafa,AENSS}. Localization was shown for a  $\Z^{d}$-ergodic random displacement model where the displacement probability distribution has a bounded density \cite{Klop93}.
  In contrast, a lot less is known about $\R^d$-ergodic random Schr\"odinger operators (random amorphous media).  There are localization results   for a class of Gaussian random potentials \cite{FLM,U,LMW}.   Localization for Poisson models where the single-site potentials are multiplied by random variables with bounded densities  has also   been studied \cite{MS,CH}. What all these results have in common is the availability of random variables with  densities which can be exploited, in an averaging procedure,  to produce an \emph{a priori} Wegner estimate at all scales (e.g.,  \cite{HM,CH,Klop95,CHM,Kir,FLM,CHN,CHKN,CHK}).

In contrast, 
for the  most natural  random Schr\"odinger operators on the continuum (cf.\  \cite[Subsection~1.1]{LGP}),   the Poisson Hamiltonian (simplest disordered amorphous medium)  and the Bernoulli-Anderson Hamiltonian (simplest disordered substitutional alloy),
until  recently there were no localization results in two or more dimensions.   The latter is an   Anderson-type Hamiltonian where the  coefficients of the single-site potentials are Bernoulli random variables.  In both cases the  random variables with  bounded densities (or at least H\"older continuous distributions \cite{CKM,St2}) are not available.

Localization for the Bernoulli-Anderson Hamiltonian has been recently proven by 
Bourgain and Kenig \cite{BK}. In this remarkable paper the Wegner estimate is established   by a multiscale analysis using  ``free sites" and a new quantitative version of unique  continuation which gives  a lower bound on eigenfunctions. 
Since their Wegner estimate  has weak probability estimates and  the underlying random variables are discrete, they also introduced a new method to prove Anderson localization from
estimates on the finite-volume resolvents given by a single-energy multiscale analysis.  The new
method does not use spectral averaging as in \cite{CH,SW}, which requires random variables with bounded densities.  It is also not an energy-interval multiscale analysis as in \cite{vDK,FMSS,Kle}, which requires better probability estimates.

The Bernoulli-Anderson Hamiltonian is the random Schr\"odinger operator $H_{\X}$  in \eqref{PoissonH} with $\X$  a  Bernoulli process on $\Z^d$ (i.e., $\X= \{j \in \Z^d; \beps_j=1\}$ with $\{\beps_j\}_{j \in \Z^d}$ independent Bernoulli random variables).  Since Poisson processes can be approximated  by appropriately defined Bernoulli processes, one might expect to prove localization for  Poisson Hamiltonians  from  the Bourgain-Kenig results  using this approximation.  This approach was indeed  used by Klopp \cite{Klop97} to study the density of states of Poisson Hamiltonians.  But localization is a much subtler phenomenon,  and such an approach turns out to be too naive.

There are  very  important differences between  the Poisson Hamiltonian and the
 Bernoulli-Anderson Hamiltonian.
While for the latter the impurities are placed on the fixed configuration $\Z^{d}$, for the former
the configuration of the impurities is random, being given by a Poisson  process  on $ \R^d$.  Moreover, unlike the Bernoulli-Poisson Hamiltonian, the Poisson Hamiltonian is not monotonic with respect to the randomness.  Another difference is that the probability space for the   Bernoulli-Anderson Hamiltonian is defined by a countable number of independent discrete (Bernoulli) random variables, but the probability space of a Poisson process is not so simple, leading to measurability questions absent in the case of the  Bernoulli-Anderson Hamiltonian. These differences are of particular importance in proving localization  as Bourgain and Kenig  required some detailed knowledge about the location of the impurities, as well as information on ``free sites'', and relied on conditional probabilities.

To prove localization for Poisson Hamiltonians, we develop  a multiscale analysis that 
exploits  the   probabilistic properties of Poisson point processes to
 control  the randomness of the configurations, and at the same time allows  the  use of the  new ideas introduced by  Bourgain and Kenig.

 \subsection{Main results}

 In this article  the single-site potential  $u$ is a  nonnegative, nonzero  $\mathrm{L}^{\infty}$-function on $\R^{d}$ with compact support, with 
 \begin{equation} \label{u}
u_{-}\chi_{\Lambda_{\delta_{-}}(0)}\le u \le u_{+}\chi_{\Lambda_{\delta_{+}}(0)}\quad \text{for some constants $u_{\pm}, \delta_{\pm}\in ]0,\infty[ $}
\end{equation}
where $\Lambda_{L}(x)$ denotes  the 
box of side $L$ centered at $x \in \R^{d}$.

We need to introduce some notation. For a given set $B$, we denote by $\chi_{B}$ its characteristic function, by  $\cP_{0}(B)$ the collection  of all countable subsets of $B$, and by  $\#B$  its cardinality.  Given $X \in  \cP_{0}(B)$ and $A\subset B$, we set $X_{A}:=X \cap A$ and 
$N_{X}(A):= \# X_{A} $.   Given a Borel set $A \subset \R^d$, we write  $\abs{A}$ for its   Lebesgue measure.  We let 
 $\Lambda_{L}(x):=x+ \left(-\frac L 2,\frac L 2\right)^{d}$  be  the 
box of side $L$ centered at $x \in \R^{d}$.  By $\Lambda$ we will always denote some box  $ \Lambda_L(x)$ , with $\Lambda_{L}$ denoting a box  of side $L$.
We set
 $\chi_{x}:= \chi_{\Lambda_{1}}(x)$,   the characteristic  function of the
 box of side $1$ centered at
 $x \in \mathbb{R}^d$. We write
$\langle x \rangle := \sqrt{1+|x|^2}$, 
$T(x) :=\scal{x}^\nu$ for some fixed $\nu>\frac d 2$. By $C_{a,b, \ldots}$, $K_{a,b, \ldots}$, etc., 
 will always denote some finite constant depending only on 
$a,b, \ldots$.

  A Poisson process on a Borel set  $B \subset \R^d$ with density (or intensity) $\varrho >0$ is a map $\X$ from a probability space  $(\Omega,\P)$ to  $\cP_{0}(B)$,
  such that  for each Borel set $A\subset B$ with $\abs{A}<\infty$ the random variable  $N_{\X}(A)$     has  Poisson distribution with mean  $\varrho |A|$,  i.e., 
 \beq
 {\P}\{N_{\X}(A)=k\}=\tfrac {(\varrho |A|)^k} {k!} \mathrm{e}^{-\varrho |A|}\quad \text{for $k=0,1,2,\dots$},
 \eeq
  and the random variables  $\{N_{\X}(A_{j})\}_{j=1}^{n}$ are independent for disjoint Borel subsets $\{A_{j}\}_{j=1}^{n}$ (e.g., \cite{King,Reiss}).
  
  The Poisson Hamiltonian $H_\X$ is an $\R^d$-ergodic family of
random self-adjoint operators.
It follows from standard results (cf.\ \cite{KM,PF})
that there exists fixed subsets of $\R$ so that the spectrum
of $H_\X$,  as well as the pure point,
absolutely continuous, and singular continuous components,
are equal to these fixed sets with probability one. It follows from our assumptions on 
the single-site potential  $u$ that  $\sigma(H_\X)=[0,+\infty[$ with probability one  \cite{KM}.

For Poisson random potentials the density $\vrho$ is a measure of the amount of disorder in the medium.  Our first result gives localization at fixed disorder at the bottom of the spectrum.

\begin{theorem}\label{thmpoisson}  Let $H_{\X}$ be a Poisson Hamiltonian on  $\L^{2}(\R^{d})$ with density $\vrho >0$.  Then
there exist $E_{0}=E_0(\vrho)>0$ and $m= m(\rho)>0$ for which  the following holds  ${\P}$-a.e.: The operator $H_\X$ has pure point spectrum  in $[0,E_0]$ with exponentially localized eigenfunctions with rate of decay $m$, i.e., if    $\phi$ is an eigenfunction of $H_\X$ 
with eigenvalue $E \in[0,E_0]$ we have
\begin{equation}\label{expdecay}
 \|\chi_x \phi\| \le C_{\X,\phi} \, e^{-m|x|}, \quad \text{for all $x \in \R^{d}$}.
\end{equation}
Moreover, there exist   $\tau>1$ and  $s\in]0,1[$ such that for  all eigenfunctions $\psi,\phi$ (possibly equal) with the same eigenvalue  $E\in[0,E_0]$ 
we have 
\begin{equation}\label{SUDEC}
\| \chi_x\psi\| \, \|\chi_y \phi\| \le C_{\X} \|T^{-1}\psi\|\|T^{-1}\phi\| \, e^{\scal{y}^\tau} e^{-|x-y|^s}, \quad \text{for all $x,y \in \Z^{d}$}.
\end{equation}
 In particular, the eigenvalues of $H_\X$ in $[0,E_0]$ have finite multiplicity,  and  $H_\X$ exhibits dynamical localization in $[0,E_0]$, that is, for any $p>0$  we have
\begin{equation}\label{dynloc}
\sup_t \| \scal{x}^p e^{-itH_\X} \chi_{[0,E_0]}(H_\X) \chi_0 \|^2_2 < \infty.
 \end{equation}
\end{theorem}

  The next theorem gives localization at high disorder in a fixed interval at the bottom of the spectrum.

\begin{theorem}\label{thmpoissonbis}  Let $H_{\X}$ be a Poisson Hamiltonian on  $\L^{2}(\R^{d})$ with density $\vrho >0$.
Given $E_0 >0$, there exist $\varrho_0=\varrho_0(E_{0})>0$ and $m=m(E_{0})>0$  such that the conclusions of Theorem~\ref{thmpoisson} hold in the interval  $[0,E_0]$
if  $\varrho>\varrho_0$ .
\end{theorem}

  Theorems~\ref{thmpoisson} and \ref{thmpoissonbis} are proved by a multiscale analysis as in \cite{B,BK}, where  the Wegner estimate, which gives control on the finite volume resolvent, is obtained by induction on the scale.  In contrast, the usual proof of localization by a multiscale analysis \cite{FS,FMSS,S,vDK,CH,FK,GK1,Kle}  uses an \emph{a priori} Wegner estimate valid  for all scales.
Exponential localization will then follow from this new single-energy multiscale analysis as in \cite[Section~7]{BK}.  The decay of  eigenfunction correlations exhibited  in (\ref{SUDEC})    follows from a detailed analysis of  \cite[Section~7]{BK} given in \cite{GKsudec2}, using ideas from  \cite{GKsudec}.  Dynamical localization and finite multiplicity of eigenvalues follow from   (\ref{SUDEC}).  That (\ref{SUDEC}) implies dynamical localization is rather immediate.  The finite multiplicity of the eigenvalues follows by estimating $\| \chi_x \chi_{\{E\}}(H_\X)\|_2^2 \|\chi_y  \chi_{\{E\}}(H_\X)\|_2^2$ from (\ref{SUDEC}) and summing over $x \in \Z^d$.

Bourgain and Kenig's methods \cite{BK} were developed for  the Bernoulli-Anderson Hamiltonian. Let $\beps_{\Z^d}=\{\beps_{\zeta}\}_{\zeta \in \Z^d}$
denote independent identically distributed
 Ber\-noulli random variables, $\beps_{\zeta}=0$ or $1$ with equal
probability. The Bernoulli-Anderson random potential is
$V(x) = \sum_{\zeta\in \Z^d}  \beps_{\zeta} u(x-\zeta)$,
and the Hamiltonian has the form (\ref{PoissonH}). To see the connection
with the Poisson Hamiltonian, let us introduce the Bernoulli-Poisson
Hamiltonian. We consider a  configuration
$Y \in \cP_{0}(\R^{d})$, and let $\beps_{Y}=\{\beps_{\zeta}\}_{\zeta \in Y}$
be the corresponding collection of  independent identically distributed Bernoulli random variables.
We define the Bernoulli-Poisson Hamiltonian
by $H_{(Y,\beps_{Y})}:= -\Delta + \sum_{\zeta\in Y}  \beps_{\zeta}
u(x-\zeta)$. In this notation,
the Bernoulli-Anderson Hamiltonian is  $H_{(\Z^{d},\beps_{\Z^{d}})}$.
If $\Y$ is a Poisson process on $\R^d$ with density
$2\varrho $, then  $\X= \{\zeta \in \Y; \, \beps_{\zeta}=1\}$ is
a Poisson process  on $ \R^d$ with density  $\varrho $, and it follows
that $H_{\X}= H_{(\Y,\beps_{\Y})}$.  Thus the Poisson Hamiltonian $H_{\X}$ 
can be rewritten as the Bernoulli-Poisson Hamiltonian  $H_{(\Y,\beps_{\Y})}$.

For  the Bernoulli-Anderson Hamiltonian the impurities are placed on the fixed configuration $\Z^{d}$, where for the the Bernoulli-Poisson Hamiltonian
the configuration of the impurities is random, being given by a Poisson  process  on $ \R^d$.   Moreover,  the probability space for the   Bernoulli-Anderson Hamiltonian is quite simple, being defined by a countable number of independent discrete (Bernoulli) random variables, but the more complicated probability space of a Poisson process leads to measurability questions absent in the case of the  Bernoulli-Anderson Hamiltonian.    We incorporate the control of the randomness of the configuration in the   multiscale analysis, ensuring  detailed knowledge about the location of the impurities, as well as information on ``free sites''.

In order to control and keep track of the random  location of the impurities,
and also handle the  measurability questions that appear for the Poisson process, we  perform a finite volume reduction in each scale as part of the multiscale analysis, which estimates the probabilities of \emph{good} boxes. 
We exploit properties of Poisson processes to construct, inside a  box $\Lambda_{L}$, a scale dependent  class of $\Lambda_{L}$-\emph{acceptable} configurations of high probability for the Poisson process $\Y$
(Definition~\ref{deflat} and Lemma~\ref{lemlat}).  We introduce an equivalence relation for  $\Lambda_{L}$-{acceptable} configurations and, 
 showing  that we can move an impurity a little bit without spoiling the goodness of boxes (Lemma~\ref{lemmovepoint}),  we conclude that  \emph{goodness} of boxes is a property of equivalence classes of acceptable configurations (Lemma~\ref{lemqgood}). Basic configurations and events in a given box are introduced in terms of these equivalence classes of acceptable configurations, and the multiscale analysis is performed for basic events.   Thus we will have a new step in the  multiscale analysis:  basic configurations and events
 in a given box will have to be rewritten in terms of  basic configurations and events in a bigger box (Lemma~\ref{lemredraw}).  The  Wegner estimate at scale $L$ is proved in Lemma~\ref{lemwegner} using \cite[Lemma~5.1$^{\pr}$]{BK}.

 Theorems~\ref{thmpoisson} and \ref{thmpoissonbis} were announced in \cite{GHK}.
Random Schr\"odinger operators with an attractive Poisson random potential, i.e., $H_{\X}=-\Delta - V_{\X}$ with $V_{\X}$ a  Poisson random potential as in this paper, so   $\sigma(H_\X)=\R$ with probability one, 
are studied in \cite{GHK2}, where we modify the methods of this paper to prove localization at low energies.

This paper is organized as follows. In Section~\ref{sectpre} we describe the construction of a Poisson process $\X$ from a marked Poisson process $(\Y,\beps_{\Y})$, and review some useful deviation estimates for Poisson random variables. Section~\ref{secfinvol} is devoted to finite volume considerations and the control of Poisson configurations:  We introduce finite volume operators,  perform the  finite volume reduction,  study the effect of changing scales, and introduce localizing events.  In Section~\ref{sectinit} we prove  \emph{a priori} finite volume estimates that give the starting hypothesis for the multiscale analysis.  Section~\ref{sectMSA}
contains the multiscale analysis for Poisson Hamiltonians.  Finally, the proofs of
Theorems~\ref{thmpoisson} and \ref{thmpoissonbis} are completed in Section~\ref{sectEND}.

\section{Preliminaries} \label{sectpre}

\subsection{Marked Poisson process}

We may assume that a  Poisson process $\X$ on
  $ \R^d$ with density  $\varrho$ is constructed from a marked Poisson process as follows:  Consider a Poisson process $\Y$ on
 $ \R^d$ with density  $2\varrho $, and to each $\zeta \in Y$
associate a Bernoulli random variable $\beps_{\zeta}$, either $0$ or $1$ with equal probability, with $\beps_{\Y}=\{\beps_{\zeta}\}_{\zeta \in \Y}$ independent random variables.  Then $(\Y,\beps_{\Y})$ is a Poisson process with density $2 \rho$ on the product space $ \R^d \times \{0,1\}$,
the \emph{marked Poisson process};  its    underlying probability space will still be denoted by $(\Omega,\P)$.   (We use the notation 
 $(Y,\eps_{Y}):=\{(\zeta,\eps_{\zeta}); \, \zeta \in Y\} \in \cP_{0}(\R^{d}\times \{0,1\})$. 
  A Poisson process on $ \R^d \times \{0,1\}$ with density $\mu>0$  is a map $\tilde{\mathbf{Z}}$ from a probability space   to  $ \cP_{0}( \R^d \times \{0,1\})$,
  such that  for each Borel set $\tilde{A}\subset  \R^d \times \{0,1\}$ with $|{\tilde{A}}|:= \frac 1 2 (|{ \{ x \in \R^{d} ; \, (x,0) \in \tilde{A}\}}|+|{ \{ x \in \R^{d} ; \, (x,1) \in \tilde{A}\}}|   )          <\infty$, the random variable  $N_{\tilde{\mathbf{Z}}}(\tilde{A})$     has  Poisson distribution with mean  $\mu |\tilde{A}|$,
  and the random variables  $\{N_{\tilde{\mathbf{Z}}}(\tilde{A}_{j})\}_{j=1}^{n}$ are independent for disjoint Borel subsets $\{\tilde{A}_{j}\}_{j=1}^{n}$.
  Define   maps $\cX,\cX^{\pr}\colon \cP_{0}( \R^d \times \{0,1\})\to  \cP_{0}( \R^d)$ by
\begin{equation}\label{XY}
\cX (\tilde{Z}):= \{ \zeta\in \R^{d} ; \, (\zeta,1) \in \tilde{Z}\} , \quad  \cX^{\pr} (\tilde{Z}):= \{ \zeta\in \R^{d} ; \, (\zeta,0) \in \tilde{Z}\},
\end{equation}
for all $\tilde{Z} \in  \cP_{0}( \R^d \times \{0,1\})$.  
Then the maps $\X,
\X^{\pr}\colon \Omega \to \cP_{0}( \R^d)$,  given 
 by 
\begin{equation}\label{XYomega}
\X:=\cX(\Y,\beps_{\Y}), \quad \X^{\pr}:=\cX^{\pr}(\Y,\beps_{\Y}),
\end{equation}
i.e., $\X(\omega)=\cX(\Y(\omega),\beps_{\Y(\omega)}(\omega))$,
$\X^{\pr}(\omega)=\cX^{\pr}(\Y(\omega),\beps_{\Y(\omega)}(\omega))$, are     Poisson processes on $ \R^d$ with density $\vrho$.  (See \cite[Section~5.2]{King}, \cite[Example~2.4.2]{Reiss}.)  In particular,  note that 
 \begin{equation}\label{NXNY}
N_{\X}(A) + N_{\X^{\pr}}(A)=  N_{\Y}(A) \quad \text{ for all Borel sets  $A \subset \R^{d}$}.
\end{equation}

If $\X$ is a    Poisson process on $\R^{d}$ with density $\vrho$, then  $\X_{A}$ is a Poisson process on $A$ with density $\vrho$  for each Borel set  $A \subset \R^{d}$,   with  $\{\X_{A_{j}}\}_{j=1}^{n}$ being  independent Poisson processes   for disjoint Borel subsets $\{A_{j}\}_{j=1}^{n}$. Similar considerations apply to $\X^{\pr}$ and to 
the marked  Poisson process   $(\Y,\beps_{\Y})$,  with $\X_{A},\X^{\pr}_{A},\Y_{A},\beps_{\Y_{A}}$ satisfying  \eqref{XYomega}.

\subsection{Poisson random variables}
For   a Poisson random variable $N$  with mean $\mu$  we have
(e.g., \cite[Eq.~(1.12)]{King})
\begin{equation}\label{Poissonk0}
 \P\{ N \ge k\}= \int_{0}^{\mu }\di \lambda  \, \frac {\lambda^{k-1}}{(k-1)!}\e^{-\lambda}, \quad \text{for $k=1,2,\ldots$},
\end{equation}
and hence also
\begin{equation}\label{Poissonk01}
 \P\{ N < k\}= \int_{\mu}^{\infty}\di \lambda  \, \frac {\lambda^{k-1}}{(k-1)!}\e^{-\lambda}, \quad \text{for $k=1,2,\ldots$}.
\end{equation}
From \eqref{Poissonk0} we get  useful upper and lower bounds:
\begin{equation}\label{Poissonk}
\frac {\mu^{k}}{k!} \e^{-\mu}< \P\{ N \ge k\}< \frac {\mu^{k}}{k!},  \quad \text{for $k=1,2,\ldots$}.
\end{equation}
When  $k > \e \mu >1$, we can use a lower bound from Stirling's formula \cite{Ro} to get 
\begin{equation}\label{Poissonklarge}
\P\{ N \ge k\}< \frac 1 {\sqrt{2\pi k}} \left(\frac {\e \mu } k\right)^{k} .
\end{equation}
  In particular, if $\e \mu >1$ and $a > \e^{2}$ we get the large deviation estimate 
\begin{equation}\label{Poissonvol}
\P\{ N \ge a \mu \}< \e^{-a \mu}.
\end{equation}
From \eqref{Poissonk01} we get
\begin{equation}\label{Poissonk012}
 \P\{ N < k\}< C_{k}\e^{-{\frac \mu 2}},\quad \text{with} \quad C_{k}=    \int_{0}^{\infty}\di \lambda  \, \frac {\lambda^{k-1}}{(k-1)!}\e^{-\frac {\lambda} 2} \quad \text{for $k=1,2,\ldots$}.
\end{equation}

\section{Finite volume and  Poisson configurations}  \label{secfinvol}
 
From now on $H_{\X}$ will always denote a Poisson Hamiltonian on  $\L^{2}(\R^{d})$ with density $\vrho >0$, as in (\ref{PoissonH})-(\ref{u}).    We recall that $(\Omega,\P)$ is the underlying probability space on which the Poisson processes $ \X$ and $\X^{\pr}$, with density $\vrho$,
and $\Y$, with density $2\vrho$, are defined, as well as the Bernoulli random variables $\beps_{\Y}$, and we have \eqref{XYomega}.  All events will be defined with respect to this probability space.  We will use the notation  $\sqcup$ for disjoint unions: $C=A\sqcup B$ means  $C=A \cup B$ with  $A \cap B =\emptyset$. 

Given  two disjoint  configurations $X,Y \in \cP_{0}(\R^{d})$ and $t_{Y}=\left\{t_{\zeta} \right\}_{\zeta \in Y} \in [0,1]^{Y}$, we set 
 \begin{equation} \label{PoissonHY}
H_{X,(Y, t_{Y})}:=  -\Delta + V_{X,(Y, t_{Y})} , \ \text{where} \ V_{X,(Y, t_{Y})}(x) := V_{X}(x) + \sum_{\zeta \in Y} t_{\zeta}u(x - \zeta).
\end{equation}
In particular, given $\eps_{Y}\in \{0,1\}^{Y}$ we have, recalling \eqref{XY}, that
  \begin{equation}\label{PoissonHYeps}
H_{X,(Y, \eps_{Y})}= H_{ X \sqcup \cX(Y,\eps_{Y})}.
\end{equation}
We also write $H_{(Y, t_{Y})}:=H_{\emptyset,(Y, t_{Y})}$ and 
\begin{equation} \label{Homega}
H_{\omega}:= H_{\X(\omega)}= H_{(\Y(\omega),\beps_{\Y(\omega)}(\omega))}.
\end{equation}

\subsection{Finite volume operators}
Finite volume operators are defined as follows:  Given   a box  $\Lambda= \Lambda_{L}(x)$  in $\R^{d}$ and a configuration $X\in \cP_{0}(\R^{d})$, we set 
\begin{align}\label{finvolH}
H_{X,\Lambda} :=-{\Delta_{\Lambda}}+ V_{X,\Lambda} \quad \text{on}   \quad \L^{2}(\Lambda),
\end{align}
where $\Delta_{\Lambda}$ is the  Laplacian on $\Lambda$ with Dirichlet boundary condition,  and  
\begin{equation}
 V_{X,\Lambda}:= \chi_{\Lambda} V_{X_{\Lambda}} \quad \text{with $V_{X_\Lambda} $ as in \eqref{PoissonV}}.  \label{finvolV}
\end{equation}
The finite volume resolvent  is 
 $R_{X,\Lambda} (z):=(H_{X,\Lambda} - z)^{-1}$.

We have 
$\Delta_{\Lambda}= \nabla_{\Lambda}\cdot\nabla_{\Lambda}$, where 
$\nabla_{\Lambda}$ is the gradient with Dirichlet boundary condition.
We sometimes identify $\L^{2}(\Lambda)$ with $\chi_{\Lambda }\L^{2}(\R^{d})$
and, when necessary,  will use subscripts $\Lambda$ and $\R^{d}$ to distinguish between the norms and inner products of $\L^{2}(\Lambda)$ and $\L^{2}(\R^{d})$.
Note that in general we do not have $ V_{X,\Lambda}=  \chi_{\Lambda} V_{X,{\Lambda^{\prime}}}$ for   $\Lambda \subset \Lambda^{\prime}$, where $\Lambda^{\prime}$ may be a finite box or $\R^{d}$. But we always have
\begin{equation}
 \chi_{\widehat{\Lambda}} V_{X,\Lambda}=  \chi_{\widehat{\Lambda}} V_{X,{\Lambda^{\prime}}}, 
\end{equation}
where
\beq    \widehat{\Lambda}=\widehat{\Lambda}_{L}(x):=\Lambda_{L-\delta_{+} }(x) \quad  \text{with $\delta_{+}$ as in \eqref{u}},  \label{lambhat}
\eeq
which suffices for the multiscale analysis.

  The multiscale analysis estimates probabilities of desired properties of finite volume resolvents at  energies  $E\in \R$. (By $L^{p\pm}$ we mean $L^{p\pm \delta}$ for some small $\delta >0$, fixed   independently of the scale.)
  
  \begin{definition} Consider an energy $E\in \R$,  a rate of decay $m>0$, and a configuration $X \in \cP_{0}(\R^{d})$.  A   box  $\Lambda_{L}$ is said to be $(X,E,m)$-good if 
  \begin{align}\label{weg}
\| R_{X,\Lambda_{L}}(E) \|& \le \e^{L^{1-}}
\intertext{and} 
\| \chi_x R_{X,\Lambda_{L}}(E) \chi_y \|& \le \e^{-m |x-y|},\quad  \text{for all  $x,y \in \Lambda_{L}$ with $ |x-y|\ge \tfrac L{10}$}. \label{good}
\end{align}
The box $\Lambda_{L}$ is $(\omega,E,m)$-good if it is $(\X(\omega),E,m)$-good.  
 \end{definition}

Note that  \cite[Lemma~2.14]{BK} requires condition \eqref{good} as stated above  for its proof.
    
But  \emph{goodness} of boxes does not suffice for the induction step in the multiscale analysis given in \cite{B,BK}, which also needs an adequate supply of \emph{free sites}  to obtain a Wegner estimate at each scale.   Given  two disjoint  configurations $X,Y \in \cP_{0}(\R^{d})$ and $t_{Y}=\left\{t_{\zeta} \right\}_{\zeta \in Y} \in [0,1]^{Y}$, we recall \eqref{PoissonHY} 
and define the corresponding finite volume operators $H_{X,(Y, t_{Y}),\Lambda}$ as in \eqref{finvolH}  and \eqref{finvolV} using $X_{\Lambda}$,  $Y_{\Lambda}$ and $t_{Y_{\Lambda}}$, i.e.,
\beq
H_{X,(Y, t_{Y}),\Lambda} := -\Delta_{\Lambda}+ V_{X,(Y, t_{Y}),\Lambda}, \quad \text{where} \quad  V_{X,(Y, t_{Y}),\Lambda}:= \chi_{\Lambda} V_{X_{\Lambda},(Y_{\Lambda}, t_{Y_{\Lambda}})},
\eeq
 with 
 $R_{X,(Y, t_{Y}),\Lambda} (z)$ being the corresponding finite volume resolvent.

\begin{definition} Consider an energy $E\in \R$,  a rate of decay $m>0$, and two   configurations $X, Y \in \cP_{0}(\R^{d})$.  A  box  $\Lambda_{L}$ is said to be $(X,Y,E,m)$-good if $X\cap Y=\emptyset$ and 
 we have   \eqref{weg} and  \eqref{good} 
with  $R_{X,(Y, t_{Y}),\Lambda_{L}} (E)$ for all $t_{Y} \in [0,1]^{Y}$.  In this case $Y$ consists of $(X,E)$-free sites for the box  $\Lambda_{L}$. \emph{(In particular, the box  $\Lambda_{L}$ is  $(X\sqcup \cX(Y,\eps_{Y}),E,m)$-good for all  $\eps_{Y} \in\{0,1\}^{Y}$.)}
\end{definition}

 \subsection{Finite volume reduction of Poisson configurations}

The multiscale analysis will require some detailed knowledge about the location of the impurities, that is, about the Poisson process configuration, as well as information on ``free sites''. To do so and also handle the  measurability questions that appear for the Poisson process we will perform a finite volume reduction as part of the multiscale analysis. The key is that we can move a Poisson point a little bit without spoiling the goodness of boxes, using the following lemma.

\begin{lemma}\label{lemmovepoint} Let $\Lambda$ be a box in $\R^{d}$,
$0\le W \in \L^{1}_{\mathrm{loc}}(\Lambda)$, $0\le w \in  \L^{\infty}(\Lambda)$ with compact support.  Given  $\zeta \in \Lambda^{(w)}=\{  \zeta \in \Lambda; \  \supp w(\cdot - \zeta) \subset \Lambda\}$, 
let
 $H_{\zeta}= - \Delta_{\Lambda} + W +w(\cdot - \zeta)$ on $\L^{2}(\Lambda)$, with  $R_{\zeta}(z)= (H_{\zeta}-z)^{-1}$ its resolvent. \\
 \noindent{\emph{\textbf{(i)}}} Suppose that for some $\zeta \in \Lambda^{(w)}$,  $E \ge 0$, and $\gamma \ge 1$  we have 
 $\norm{R_{\zeta}(E)} \le  \gamma $, and let 
\begin{equation}\label{mv0}
0 < \eta \le  \min \left\{\left(4\sqrt{1 +E}
\norm{w}_{\infty } \gamma \right)^{-2},\tfrac 1 4\right\}.
\end{equation}
Then for all $\zeta^{\prime} \in \Lambda^{(w)}$ with $\abs{\zeta^{\prime} - \zeta}\le \eta$  we have 
  \begin{align}\label{mv1}
\| R_{\zeta^{\prime}}(E) \|& \le \e^{\sqrt{\eta}}\gamma
\intertext{and} 
\| \chi_x R_{\zeta^{\prime}}(E) \chi_y \|& \le\| \chi_x R_{\zeta}(E) \chi_y \| + \sqrt{\eta}\,\gamma,
\quad   \text{for all  $x,y \in \Lambda$}. \label{mv2}
\end{align} 
 \noindent{\emph{\textbf{(ii)}}} Suppose that for some $\zeta \in \Lambda^{(w)}$,  $E \ge 0$, and $\beta \ge 2$  we have  $\dist (E, \sigma(H_{\zeta})) \le \beta^{-1}$, i.e.,
 $\norm{R_{\zeta}(E)} \ge  \beta$, and let  $\eta$ be as in \eqref{mv0} with $\beta$ substituted for $\gamma$.
Then for all $\zeta^{\prime} \in \Lambda^{(w)}$ with $\abs{\zeta^{\prime} - \zeta}\le \eta$  we have 
\beq \label{etabeta}
\| R_{\zeta^{\prime}}(E) \| \ge \e^{- \sqrt{\eta} } \beta, \quad \text{i.e.,}\quad
\dist (E, \sigma(H_{\zeta^{\pr}}))  \le  \e^{ \sqrt{\eta}  }\beta^{-1}.
\eeq
\end{lemma}

\begin{proof}
We set  $R=R_{\zeta}(E)$, $R^{\prime}=R_{\zeta^{\prime}}(E)$, $u= w(\cdot - \zeta)$,  $u^{\prime}= w(\cdot - \zeta^{\prime})$, and  $\xi= \zeta^{\prime} - \zeta$ with
$\abs{\xi} \le \eta$. 
We let $U(a)$  denote translation by $a$ in $L^{2}(\R^{d})$:  $(U(a)\vphi)(x)= \vphi(x -a)$,
and  pick $\phi \in C_{c}^{\infty}({\Lambda})$ such that $ 0\le \phi \le 1$ and $\phi\equiv 1$ in some open subset of $\Lambda$ which contains the supports of $u$ and $u^{\prime}$.  It follows from the resolvent identity that
\begin{align}\nonumber
& \norm{R^{\prime}}_{\Lambda}- \norm{R}_{\Lambda} \le \left\|R^{\prime} (u^{\prime}-u)R \right\|_{\Lambda}=\left\|\chi_{\Lambda}R^{\prime}\phi (u^{\prime}-u)\phi R\chi_{\Lambda} \right\|_{\R^{d}}\\ \label{mv3}
& \quad   = \left\|\chi_{\Lambda}R^{\prime} \phi (U(\xi)uU(\xi)^{*}- u)\phi R\chi_{\Lambda}\right\|_{\R^{d}} \\
& \quad \nonumber
 \le \left\|\chi_{\Lambda}R^{\prime} \phi (U(\xi)-1)u U(\xi)^{*}\phi R\chi_{\Lambda} \right\|_{\R^{d}}  + \left\|\chi_{\Lambda}R^{\prime}\phi u(U(\xi)^{*}-1)\phi R\chi_{\Lambda} \right\|_{\R^{d}} \\
 & \quad \nonumber \le 
\eta \left(  \norm{u \nabla \phi R^{\prime }\chi_{\Lambda} }_{\R^{d}} 
 \norm{\phi R\chi_{\Lambda} }_{\R^{d}}  +\norm{\phi R^{\prime}\chi_{\Lambda} }_{\R^{d}} \norm{u \nabla \phi R\chi_{\Lambda} }_{\R^{d}} \right)\\
 & \quad \nonumber =
\eta \left( \norm{u \nabla_{\Lambda}\phi R^{\prime} }_{\Lambda}  
 \norm{\phi R }_{\Lambda}  + \norm{\phi R^{\prime} }_{ \Lambda}
 \norm{u\nabla_{\Lambda}\phi R }_{\Lambda} \right)\\
 & \quad \nonumber \le
\eta\norm{u}_{\infty } \left( \norm{ \nabla_{\Lambda} R^{\prime} }_{\Lambda}  
 \norm{ R }_{\Lambda}  + \norm{ R^{\prime} }_{ \Lambda}
 \norm{\nabla_{\Lambda} R }_{\Lambda} \right)\\
  & \quad \nonumber \le 2\sqrt{1 +E}
\norm{u}_{\infty }\eta  \max\{\norm{ R }_{\Lambda}, 1\}   \max\{\norm{ R^{\pr} }_{\Lambda}, 1\} , 
 \end{align}
 where we used \begin{equation}
 \norm{\nabla_{\Lambda} R^{\sharp} }_{\Lambda}^{2} \le  \norm{ R^{\sharp} }_{\Lambda}
 + E \norm{ R^{\sharp}  }_{\Lambda}^{2}\le (1 +E) \max\{\norm{R^{\sharp} }_{\Lambda}^{2}, 1\} \;\; \text{for} \;\; R^{\sharp} =R, R^{\pr}.
\end{equation}

To prove part (i), if  $\norm{R}_{\Lambda} \le  \gamma $ with $\gamma \ge 1$, it follows from 
\eqref{mv3} and  \eqref{mv0} that
 \beq \label{mv56}
 \norm{ R^{\prime} }_{ \Lambda} - \norm{ R }_{ \Lambda}\le\left\|R^{\prime} (u^{\prime}-u)R \right\|_{\Lambda}\le
  \tfrac 12  \sqrt{\eta} \max\{\norm{ R^{\pr} }_{\Lambda}, 1\}.
\eeq
To prove  \eqref{mv1}, we may  assume that  $\|  R^{\prime} \|_{\Lambda} \ge 1$, since otherwise the result is trivial. 
The estimate \eqref{mv1} now follows immediately from \eqref{mv56} and  \eqref{mv0}. 
Using the resolvent identity,     \eqref{mv56}, \eqref{mv1}, and $\tfrac 1 2 \e^{\tfrac 1 2}< 1$ we get \eqref{mv2}.

Part (ii) follows from part (i) as follows.   Let  $\beta \ge 2$
and suppose \eqref{etabeta} does not hold, i.e.,  $ \norm{ R^{\prime} }_{ \Lambda} < \e^{- \sqrt{\eta} } \beta$.  Since $\e^{- \sqrt{\eta} } \beta \ge \e^{- \frac 1 2 } 2> 1$, 
 we may   apply  \eqref{mv1} to get a contradiction to $ \norm{ R }_{ \Lambda}\ge \beta$, namely
$
\norm{R}_{\Lambda} <  \e^{ \sqrt{\eta} } \left(\e^{- \sqrt{\eta} } \beta\right)=\beta.
$
 \end{proof}

Lemma~\ref{lemmovepoint} lets us move one Poisson point a little bit, namely by $\eta$,  and maintain good bounds on the resolvent.  Since we will want to preserve  the ``goodness'' of the box $\Lambda=\Lambda_{L}$,  we will use Lemma~\ref{lemmovepoint} with $\gamma=\e^{L^{1-}}$ (as in \eqref{weg}), and take $\eta \ll \e^{-L}  $. To fix ideas we set $\eta=\e^{-L^{10^{6}d}}$.
  To move \emph{all} Poisson points in $\Lambda_{L}$ we will need to control the number of Poisson points in the box.   Moreover, we will have to  know  the location of these  Poisson points with good precision.  That this can be done at very little cost in probability is the subject of the next lemma. 
  
  \begin{definition}  \label{deflat}   Let  $\eta_{L}:=\e^{-L^{10^{6}d}}$ for $L>0$.  Given a box $\Lambda=\Lambda_{L}(x)$,  set
  \begin{equation}
   \J_{\Lambda}:= \{ j \in x +\eta_{L}\Z^{d}; \ \Lambda_{\eta_{L}}(j) \subset \Lambda\}.
\end{equation}
A configuration $X \in \cP_{0}(\R^{d})$ is said to be $\Lambda$-acceptable if
  \begin{gather}\label{totalN}
 N_{X}(\Lambda) < 16 \vrho L^{d},\\
  N_{X}(\Lambda_{\eta_{L}}(j ))\le 1,  \quad \text{for all} \quad j \in \J_{\Lambda},\label{tinyN}\\
     \intertext{and}
     \label{bryN}
   N_{X}(\Lambda\backslash \sqcup_{j \in  \J_{\Lambda} }\Lambda_{\eta_{L}(1-\eta_{L})}(j))=0;
 \end{gather}
it is $\Lambda$-acceptable$\,^{\prime}$ if it  satisfies \eqref{totalN},\eqref{tinyN}, and the less restrictive
 \beq  \label{bryN2}
   N_{X}(\Lambda\backslash \sqcup_{j \in  \J_{\Lambda} }\Lambda_{\eta_{L}}(j))=0.
 \eeq
We set
\begin{align}
 \cQ_{\Lambda}^{(0)}:&=\{X \in \cP_{0}(\R^{d}); \, \text{$X$ is  $\Lambda$-acceptable}\},\\
  \cQ_{\Lambda}^{(0\prime)}:&=\{X \in \cP_{0}(\R^{d}); \, \text{$X$ is  $\Lambda$-acceptable$\,^{\prime}$}\},
\end{align}
 and consider the event (recall that $\Y$ is the Poisson process with density $2\vrho$)
 \beq \label{defOmega0}
\Omega_{\Lambda}^{(0)}:=\{\Y \in \cQ_{\Lambda}^{(0)}\} .
\eeq
\end{definition}

Note that  $ \Omega_{\Lambda}^{(0)} \subset \{\X\in \cQ_{\Lambda}^{(0)}\}$ in view of \eqref{NXNY} and  $\cQ_{\Lambda}^{(0)} \subset \cQ_{\Lambda}^{(0\prime)}$. We require condition \eqref{bryN} for acceptable configurations  to avoid ambiguities when changing scales (cf. Lemma~\ref{lemredraw}), but we will then need Lemma~\ref{lemqgood} for acceptable$\,^{\prime}$ configurations. 

We now impose a condition on $\vrho$ and $L$ that will be always satisfied when we do the multiscale analysis: 
\begin{equation}
L^{-(0+) }\le \vrho\le \e^{L^{d}}. \label{vrhoL}
\end{equation} 
From now on we assume \eqref{vrhoL}.

\begin{lemma}  \label{lemlat} There exists a scale $\overline{L}=\overline{L}(d) < \infty$, such that if $L\ge \overline{L} $ 
we have
 \begin{equation}
\P \{\Omega_{\Lambda_{L}}^{(0)}\} \ge 1 - \e^{- L^{d-}} .
 \label{Omega0}
\end{equation}
\end{lemma}

\begin{proof}   Using \eqref{Poissonvol} and  \eqref{Poissonk}
we get
 \begin{equation}
\P \{\Omega_{\Lambda_{L}}^{(0)}\} \ge 1 - \e^{-16 \vrho L^{d}} - 4d \vrho  (L^{d-1}+L^{d})\eta_{L} -  2 \vrho^{2} L^{d} \eta_{L}^{d},
\end{equation}
and hence \eqref{Omega0} follows for large $L$ using \eqref{vrhoL}.
\end{proof}

Lemma~ \ref{lemlat} tells us that  inside the box $\Lambda$, outside an event of  negligible probability in the multiscale analysis, we only need to consider $\Lambda$-acceptable configurations of the Poisson process $\Y$.

Given a box  $\Lambda=\Lambda_{L}(x)$, we define an \emph{equivalence relation for configurations} by
\beq
X\overset{\Lambda}{\sim} Z \Longleftrightarrow  N_{X}(\Lambda_{\eta_{L}}(j ))=N_{Z}(\Lambda_{\eta_{L}}(j )) \quad \text{for all} \quad j \in \J_{\Lambda}\label{equivrel}.
\eeq
This induces an equivalence relation  in both $\cQ_{\Lambda}^{(0\prime)} $ and  $\cQ_{\Lambda}^{(0)} $;  the equivalence class of $X$ in $\cQ_{\Lambda}^{(0\prime)} $ will be denoted by $[X]_{\Lambda}^{\prime}$. If $X \in \cQ_{\Lambda}^{(0)}$, then  $[X]_{\Lambda}=[X]_{\Lambda}^{\prime}\cap \cQ_{\Lambda}^{(0)} $ is its  equivalence
class in $\cQ_{\Lambda}^{(0)}$. 
 Note that $[X]_{\Lambda}^{\prime}=[X_{\Lambda}]_{\Lambda}^{\prime}$. We also
write\beq  \label{[A]}
 [A]_{\Lambda}: = \bigcup_{X \in A} [X]_{\Lambda} \quad \text{for subsets}\quad  A \subset \cQ_{\Lambda}^{(0)}.
\eeq

The following lemma is  an immediate consequence of Lemma~~\ref{lemmovepoint}(i); it tells us that ``goodness'' of boxes is a property of equivalence classes of acceptable$\,^{\prime}$ configurations: changing configurations inside an equivalence class takes
good boxes  into just-as-good (jgood) boxes.

\begin{lemma} \label{lemqgood}
 Fix $E_{0}>0$ and consider an energy $E \in [0,E_{0}]$.  Suppose the box $\Lambda=\Lambda_{L}$ (with $L$ large)
is $(X,E,m)$-good for some $X \in  \cQ_{\Lambda_{L}}^{(0\prime)}$.  Then for all $Z\in  [X]_{\Lambda}^{\prime}$  the box $\Lambda$
is   $(Z,E,m)$-jgood (for  just-as-good), that is,
\begin{align}\label{qweg}
\| R_{Z,\Lambda}(E) \|& \le \e^{L^{1-} +\eta_{L}^{\frac 1 4}}\sim 
 \e^{L^{1-} }
\intertext{and} 
\| \chi_x R_{Z,\Lambda}(E) \chi_y \|& \le \e^{-m |x-y|} +
\eta_{L}^{\frac 1 4}\sim  \e^{-m |x-y|}, \; \; \text{for $x,y \in \Lambda$ with $ |x-y|\ge \tfrac L {10}$}. \label{qgood}
\end{align}
Moreover, if  $X,Z,X\sqcup Z\in  \cQ_{\Lambda}^{(0\prime)}$ and the box $\Lambda$
is $(X,Z,E,m)$-good, then for all $X_{1}\in  [X]_{\Lambda}^{\prime}$ and $Z_{1}\in  [Z]_{\Lambda}^{\prime}$   we have
$X_{1}\sqcup Z_{1}  \in [X\sqcup Z]_{\Lambda}^{\prime}$, and
 the box $\Lambda$
is  $(X_{1},Z_{1},E,m)$-jgood  as in \eqref{qweg} and \eqref{qgood}.
\end{lemma}

\begin{proof}
Lemma~~\ref{lemmovepoint}(i) gives
\begin{align}\label{qweg5}
\| R_{X^{\prime},\Lambda}(E) \|& \le \e^{L^{1-} +16 \vrho L^{d} \sqrt{\eta_{L}}},
\intertext{and, for all  $x,y \in \Lambda$ with $ |x-y|\ge \tfrac L {10}$,} 
\| \chi_x R_{X^{\prime},\Lambda}(E) \chi_y \|& \le \e^{-m |x-y|} +
16 \vrho L^{d}\sqrt{\eta_{L}} \, \e^{L^{1-} +16 \vrho L^{d} \sqrt{\eta_{L}}}. \label{qgood5}
\end{align}
Using \eqref{vrhoL}, we get \eqref{qweg} and \eqref{qgood} for large $L$.

The remaining statement is immediate.
\end{proof}

\begin{remark}  Proceeding as in Lemma~\ref{lemqgood}, we find that changing configurations inside an equivalence class takes jgood boxes  into what we may call just-as-just-as-good (jjgood) boxes, and so on.  Since we will only carry this procedure a bounded number of times, the bound independent of the scale, we will simply call them all  jgood boxes.
\end{remark}

Similarly, we get the following consequence of Lemma~~\ref{lemmovepoint}(ii).
\begin{lemma} \label{lemqgood7}
  Fix $E_{0}>0$ and consider an energy $E \in [0,E_{0}]$ and a box $\Lambda=\Lambda_{L}$ (with $L$ large).  
 Suppose $\dist (E, \sigma(H_{X,\Lambda}))  \le \tau_{L} $  for some $X \in  \cQ_{\Lambda_{L}}^{(0\prime)}$, where $  \sqrt{\eta_{L}}\ll  \tau_{L} < \frac 1 2$. Then
 \beq
\dist (E, \sigma(H_{Y,\Lambda}))  \le \e^{\eta_{L}^{\frac 1 4}} \tau_{L}, \quad \text{for all}\quad Y\in  [X]_{\Lambda}^{\prime}.
\eeq
\end{lemma}

 In view of \eqref{totalN}-\eqref{tinyN} we  have
\beq  \label{eqclasses}
\cQ_{\Lambda}^{(0)}\slash \overset{\Lambda}{\sim}\  = \{[J]_{\Lambda};\, J\in  \cJ_{\Lambda}\}, \quad \text{where} \quad \cJ_{\Lambda}:=\{ J \subset     \J_{\Lambda}; \,
\#J <  16 \vrho L^{d}\},
\eeq
and we can write $\cQ_{\Lambda}^{(0)}$ and  $\Omega_{\Lambda}^{(0)}$ as 
\beq
\cQ_{\Lambda}^{(0)}=\bigsqcup_{J\in  \cJ_{\Lambda}}[J]_{\Lambda} \quad\text{and} \quad \Omega_{\Lambda}^{(0)}=\bigsqcup_{J\in  \cJ_{\Lambda}}\{ \Y \in[J]_{\Lambda}\}.
\eeq

 \subsection{Basic events}
The multiscale analysis will require ``free sites'' and sub-events of $\{ \Y \in[J]_{\Lambda}\}$.

\begin{definition} Given $\Lambda=\Lambda_{L}(x)$,  a $\Lambda$-bconfset (basic configuration set) is a subset of $ \cQ_{\Lambda}^{(0)}$ of the form
\beq \label{cylinderset}
C_{\Lambda,B,S}:=\bigsqcup_{\eps_{S}\in \{0,1\}^{S}}[B \cup \cX(S,\eps_{S})]_{\Lambda} = \bigsqcup_{S^{\prime}\subset S} [B \cup S^{\prime}]_{\Lambda},
\eeq
 where we always implicitly assume  $B \sqcup S\in  \cJ_{\Lambda}$. 
$C_{\Lambda,B,S}$ is a $\Lambda$-dense bconfset if
  $S$ satisfies the density condition (cf. \eqref{lambhat})
  \begin{equation}  \label{densitycond}
\# (S\cap {\widehat{\Lambda}_{L^{1-}}} )\ge L^{d-}, \quad \text{for all
boxes}\quad   \Lambda_{L^{1-}}\subset \Lambda_{L}.
\end{equation}
We also set
\beq  \label{S=empty}
C_{\Lambda,B}:=C_{\Lambda,B,\emptyset}=[B]_{\Lambda} .
\eeq
\end{definition}

\begin{definition} Given $\Lambda=\Lambda_{L}(x)$,  a $\Lambda$-bevent (basic event) is  a  subset of  $ \Omega_{\Lambda}^{(0)}$ of the form
\beq \label{bevent}
\cC_{\Lambda,B,B^{\pr},S}:= \{\Y \in [B \sqcup B^{\pr}\sqcup S]_{\Lambda}\}\cap
\{\X \in C_{\Lambda,B,S}\}  \cap \{\X^{\pr} \in C_{\Lambda,B^{\pr},S}\},
\eeq
 where we always implicitly assume  $B \sqcup B^{\pr}\sqcup S\in  \cJ_{\Lambda}$. In other words, the $\Lambda$-bevent
$\cC_{\Lambda,B,B^{\pr},S}$  consists of all  $\omega \in \Omega_{\Lambda}^{(0)}$  satisfying
 \begin{align}
\begin{array}{ccl}
N_{\X(\omega)}(\Lambda_{\eta_{L}}(j ))=1 & \text{if} & j\in B ,\\
N_{\X^{\pr}(\omega)}(\Lambda_{\eta_{L}}(j ))=1 & \text{if} & j\in B^{\pr} ,\\
N_{\Y(\omega)}(\Lambda_{\eta_{L}}(j ))=1 & \text{if} & j\in S, \\
N_{\Y(\omega)}(\Lambda_{\eta_{L}}(j ))=0 & \text{if} & j\in  \J_{\Lambda}\backslash (B\sqcup B^{\pr}\sqcup S).
\end{array}
\end{align}
$\cC_{\Lambda,B,B^{\pr},S}$ is a $\Lambda$-dense bevent if
  $S$ satisfies the density condition \eqref{densitycond}. In addition, we set
\beq  \label{S=empty2}
\cC_{\Lambda,B,B^{\pr}}:=\cC_{\Lambda,B,B^{\pr},\emptyset}=\{\X \in C_{\Lambda,B}\}  \cap \{\X^{\pr} \in C_{\Lambda,B^{\pr}}\}.
\eeq
\end{definition}

The number of possible bconfsets and bevents in a given box is always finite. 
We always have 
\beq
\cC_{\Lambda,B,B^{\pr}, S}\subset \{\X \in  C_{\Lambda,B,S}\}\cap \Omega_{\Lambda}^{(0)},
\eeq
\beq
\cC_{\Lambda,B,B^{\pr},S}\subset
\cC_{\Lambda,\emptyset,\emptyset,B\sqcup B^{\pr}\sqcup S}= \{ \Y \in [B\sqcup B^{\pr}\sqcup S]_{\Lambda}\}.
\eeq
 Note also that it follows from \eqref{defOmega0}, \eqref{eqclasses} and \eqref{S=empty2} that
\beq  \label{unionOmega0}
\Omega_{\Lambda}^{(0)}= \bigsqcup_{\{(B,B^{\pr}); \, B\sqcup B^{\pr}\in\cJ_{\Lambda}\} } \cC_{\Lambda,B,B^{\pr}}
\eeq
Moreover,  for each $S_{1}\subset S$ we  have 
\begin{align} \label{cylinderexp}
C_{\Lambda,B,S}&= \bigsqcup_{S_{2}\subset  S_{1}} C_{\Lambda,B \sqcup S_{2},S\setminus S_{1}},\\
\cC_{\Lambda,B,B^{\pr},S}&= \bigsqcup_{S_{2}\subset  S_{1}} \cC_{\Lambda,B \sqcup S_{2}, B^{\pr}\sqcup (S_{1}\setminus S_{2}) ,S\setminus S_{1}}. \label{cylinderexp2}
\end{align}

In view of Lemma~\ref{lemqgood}, we make the following definition.

\begin{definition} \label{defadapted}
 Consider an energy $E \in \R$, $m>0$, and a box 
  $\Lambda=\Lambda_{L}(x)$.  
The $\Lambda$-bevent $\cC_{\Lambda,B,B^{\pr},S}$ and the $\Lambda$-bconfset $C_{\Lambda,B,S}$  are  $(\Lambda,E,m)$-good if the
box $\Lambda$ is $(B,S,E,m)$-good.   \emph{(Note that $\Lambda$ is then  $(\omega,E,m)$-jgood for every $\omega \in \cC_{\Lambda,B,B^{\pr},S}$.)}  Those
$(\Lambda,E,m)$-good bevents and bconfsets that are also $\Lambda$-dense will be called  $(\Lambda,E,m)$-adapted.     
\end{definition}

\subsection{Changing scales}
Since the finite volume reduction is scale dependent, it introduces  new  considerations in the multiscale analysis for Poisson Hamiltonians.
Given $\Lambda_{\ell}\subset \Lambda$,  the multiscale analysis will require us to redraw  $\Lambda_{\ell}$-bevents and bconfsets in terms of  $(\Lambda,\Lambda_{\ell})$-bevents and bconfsets as follows.

\begin{definition}
Given $\Lambda_{\ell}\subset \Lambda$, a configuration $J \in \cJ_{\Lambda}$
 is called $\Lambda_{\ell}$-compatible if
\beq
J\cap \Lambda_{\ell} \in \cJ_{\Lambda}^{\Lambda_{\ell}} :=\bigsqcup_{A \in \cJ_{\Lambda_{\ell}}} \J_{\Lambda}(A) \subset \cJ_{\Lambda}, 
\eeq
where
\beq
\J_{\Lambda}(A):= \{ J \subset \J_{\Lambda}\cap \Lambda_{\ell}; \, J \in [A]_{\Lambda_{\ell}}\} \quad \text{for} \quad  A \subset  \J_{\Lambda_{\ell}}.
\eeq
 If  $B\sqcup S$  is $\Lambda_{\ell}$-compatible, the $\Lambda$-bconfset $ C_{\Lambda,B,S}$ is  also called   $\Lambda_{\ell}$-compatible, and we define
the   $(\Lambda,\Lambda_{\ell})$-bconfset
 \beq
  C_{\Lambda,B,S}^{\Lambda_{\ell}}:=\{X \in \cP_{0}(\R^{d});\, X_{\Lambda_{\ell} } \in C_{\Lambda,B\cap \Lambda_{\ell},S\cap \Lambda_{\ell}}\}
  \subset \cQ_{\Lambda_{\ell}}^{(0\pr)} .\label{cylinderbox2}
 \eeq
  If  $B\sqcup B^{\pr}\sqcup S$   is $\Lambda_{\ell}$-compatible,
 the $\Lambda$-bevent  $ \cC_{\Lambda,B,B^{\pr},S}$  is  also called   $\Lambda_{\ell}$-compatible, and we  define the   $(\Lambda,\Lambda_{\ell})$-bevent   
\beq
 \cC_{\Lambda,B,B^{\pr},S}^{\Lambda_{\ell}}:= \{\Y_{\Lambda_{\ell}} \in [(B \sqcup B^{\pr}\sqcup S)\cap\Lambda_{\ell}]_{\Lambda}\}\cap
\{\X_{\Lambda_{\ell}} \in C_{\Lambda,B,S}^{\Lambda_{\ell}}\}  \cap \{\X^{\pr}_{\Lambda_{\ell}} \in C_{\Lambda,B^{\pr},S}^{\Lambda_{\ell}}\}. \label{cylinderLL2}
\eeq
Moreover,  we say that a $\Lambda_{\ell}$-compatible $\Lambda$-bconfset  $ C_{\Lambda,B,S}$ or   a  $\Lambda$-bevent  $ \cC_{\Lambda,B,B^{\pr},S}$ 
is $(\Lambda,\Lambda_{\ell})$-dense  if  $S\cap \Lambda_{\ell} $ satisfies the density condition \eqref{densitycond} in $\Lambda_{\ell}$;  $(\Lambda,\Lambda_{\ell},E,m)$-jgood  if   the box $\Lambda_{\ell}$ is $(B,S,E,m)$-jgood; 
 $(\Lambda,\Lambda_{\ell},E,m)$-adapted  if both $(\Lambda,\Lambda_{\ell})$-dense and  $(\Lambda,\Lambda_{\ell},E,m)$-jgood. \emph{ (Note that whenever we define a property of a $\Lambda$-bconfset or bevent on a subbox $\Lambda_{\ell}\subset \Lambda$  we will always implicitly assume  $\Lambda_{\ell}$-compatibility.)}  
\end{definition}

\begin{lemma}  \label{lemredraw} Let $\Lambda_{\ell}\subset \Lambda$. Then for all  $\Lambda_{\ell}$-bconfsets  $ C_{\Lambda_{\ell},B,S}$ and    $\Lambda_{\ell}$-bevents  $ \cC_{\Lambda_{\ell},B,B^{\pr},S}$  we have
\begin{align} \label{cylinderbox1}
 C_{\Lambda_{\ell},B,S} \cap \cQ_{\Lambda}^{(0)}& \subset 
\bigcup_{B_{1}\in \J_{\Lambda}(B),\, S_{1}\in \J_{\Lambda}(S)}
C_{\Lambda,B_{1},S_{1}}^{\Lambda_{\ell}}\, , \\
 \label{cylinderLL1}
 \cC_{\Lambda_{\ell},B,B^{\pr},S} \cap \Omega_{\Lambda}^{(0)} &\subset \bigsqcup_{B_{1}\in \J_{\Lambda}(B),B_{1}^{\pr}\in \J_{\Lambda}(B^{\pr}),  S_{1}\in \J_{\Lambda}(S)}
\cC_{\Lambda,B_{1},B^{\pr}_{1},S_{1}}^{\Lambda_{\ell}}.
\end{align} 
Moreover, if $C_{\Lambda_{\ell},B,S}$ or $ \cC_{\Lambda_{\ell},B,B^{\pr},S}$  is   $\Lambda_{\ell}$-dense, or  $(\Lambda_{\ell},E,m)$-jgood,  or  $(\Lambda_{\ell},E,m)$-adapted, then
  then  each $C_{\Lambda,B_{1},S_{1}}^{\Lambda_{\ell}}$ or  $\cC_{\Lambda,B_{1},B^{\pr}_{1},S_{1}}^{\Lambda_{\ell}}$ is    $(\Lambda,\Lambda_{\ell})$-dense, or $(\Lambda,\Lambda_{\ell},E,m)$-jgood, or $(\Lambda,\Lambda_{\ell},E,m)$-adapted.
 \end{lemma}

\begin{proof}
If  $C_{\Lambda_{\ell},B,S}$ is a $\Lambda_{\ell}$-bconfset, then $\{C^{\Lambda_{\ell}}_{\Lambda,B_{1},S_{1}}\}_{B_{1}\in \J_{\Lambda}(B),\, S_{1}\in \J_{\Lambda}(S)}$ form a collection of (not necessarily disjoint)  $(\Lambda,\Lambda_{\ell})$-bconfsets, and   we have  \eqref{cylinderbox1}.  The same argument yields \eqref{cylinderLL1}, but now the  $(\Lambda,\Lambda_{\ell})$-bevents are disjoint. (There are no ambiguities since $\eta_{L} \ll\sqrt{ \eta_{\ell}}$ and we have condition \eqref{bryN} at both scales.) The rest follows, using also Lemma~\ref{lemqgood}.
\end{proof}

\subsection{Localizing events} 

\begin{definition}  Consider an energy $E \in \R$, a rate of decay  $m>0$, and a box   $\Lambda$.  We call $\Omega_{\Lambda}$ a  $(\Lambda,E,m)$-localized event if
there exist  disjoint  $(\Lambda,E,m)$-adapted  bevents
	$\{\cC_{\Lambda,B_{i},B_{i}^{\pr},S_{i}}\}_{i=1,2,\ldots,I}$  such that 
\beq \label{adapted3}
\Omega_{\Lambda}= \bigsqcup_{i=1}^{I} \cC_{\Lambda,B_{i},B_{i}^{\pr},S_{i}} .
\eeq
\end{definition}

If  $\Omega_{\Lambda}$ is a  $(\Lambda,E,m)$-localized event, note  that $\Omega_{\Lambda} \subset \Omega_{\Lambda}^{(0)}$ by its definition, and hence, recalling \eqref{cylinderexp2}  and   \eqref{S=empty2} , we can rewrite $\Omega_{\Lambda} $ in the form
 \beq \label{goodcylinders}
\Omega_{\Lambda}= \bigsqcup_{j=1}^{J} \cC_{\Lambda,A_{j},A_{j}^{\pr}},
\eeq
where the $\{\cC_{\Lambda,A_{j},A_{j}^{\pr}}\}_{j=1,2,\ldots,J}$ are disjoint $(\Lambda,E,m)$-good  bevents.

We will need $(\Lambda,E,m)$-localized events of scale appropriate probability.

\begin{definition}  Fix $p>0$. Given an energy $E \in \R$ and  a rate of decay  $m>0$,  a  scale $L$ 
 is $(E,m)$-localizing if for some box  $\Lambda=\Lambda_{L}$ (and hence for all) 
 we have a $(\Lambda,E,m)$-localized event $\Omega_{\Lambda}$ such that 
\beq \label{PEL0}
\P\{\Omega_{\Lambda}\} > 1- L^{-p}.
\eeq
\end{definition}

In Section~\ref{sectEND} we will also need ``just localizing'' events and scales.
 
 \begin{definition}\label{jlocalizing}  Consider an energy $E \in \R$,  a rate of decay  $m>0$, and a box   $\Lambda$.  We call  $\Omega_{\Lambda}$ a  $(\Lambda,E,m)$-jlocalized event if
there exist  disjoint  $(\Lambda,E,m)$-good bevents
	$\{\cC_{\Lambda,A_{j},A_{j}^{\pr}}\}_{j=1,2,\ldots,J}$  such that 
\beq \label{adapted3j}
\Omega_{\Lambda}= \bigsqcup_{j=1}^{J}\cC_{\Lambda,A_{j},A_{j}^{\pr}} .
\eeq
A  scale $L$ 
 is   $(E,m)$-jlocalizing if for some box  $\Lambda=\Lambda_{L}$ (and hence for all) 
 we have a $(\Lambda,E,m)$-jlocalized event $\Omega_{\Lambda}$ such that 
\beq \label{PEL0j}
\P\{\Omega_{\Lambda}\} > 1- L^{-p}.
\eeq
 \end{definition}
 
 An  $(E,m)$-localizing scale $L$  is  $(E,m)$-jlocalizing in view of \eqref{goodcylinders}.

 \section{``A priori'' finite volume estimates}  \label{sectinit}

Given an energy $E$, to start the multiscale analysis we will need, as in \cite{B,BK},  an {\it a priori}   estimate on the probability that a box $\Lambda_{L}$ is good with an adequate supply of free sites,
for some sufficiently large scale $L$. The multiscale analysis will then show that such a probabilistic estimate also holds at all large scales.

\subsection{Fixed disorder}

\begin{proposition}\label{proppoisson}  Let $H_{\X}$ be a Poisson Hamiltonian on  $\L^{2}(\R^{d})$ with density $\vrho >0$, and  fix $p>0$.  Then  there exist   a constant  $C_{u}>0$
and a scale  $\overline{L}_{0}=\overline{L}_{0}(d,u,\vrho,p) < \infty$,  such that for all scales $L\ge\overline{ L}_{0}$ we have \eqref{vrhoL}, and, setting  
\beq  \label{EmL}
{\delta_{L}}=1 +( (p+d+1){ \vrho}^{-1} \log L)^{\frac1d}, \quad E_{L}=  C_{u}{{\delta_{L}}^{-2(d+1)}}, \quad \text{and} \quad m_{L}= \tfrac 1 2 \sqrt{E_{L}},
\eeq
   the  scale $L$  is $(E,m_{L})$-localizing for all energies $E \in [0,E_{L}]$.
\end{proposition}

The proof will be based on the following lemma.

\begin{lemma}\label{lempoisson}  Let $H_{X}$ be a Hamiltonian as in  \eqref{PoissonH}-\eqref{u}.  Given  $\delta_{0} > 0$ and  $L> \delta_{0}+ \delta_{+}$, let $\Lambda=\Lambda_{L}(x)$  and set
  \begin{equation}\label{setJ}
   J:=\{j \in x + {\delta_{0}}\Z^d \cap\Lambda; \,  \Lambda_{{\delta_{L}}}(j)\subset \widehat{\Lambda})\},  \quad  J_{e}:=J \cap (x + 2 \delta_0\Z^d ).
  \end{equation}
  Then  there 
 exist  constants  $C_{u}>0$ and $\tilde{\delta}_{u}\ge \delta_{-}$, such that if $\delta_{0}>\tilde{\delta}_{u}$, then 
     for all $ X ,Y\in \cP_{0}(\R^{d})$ and  $t_{Y}\in [0,1]^{Y}$, such that $X \cap Y=\emptyset$ and
  \beq  \label{XinJe}
   N_{X}(\Lambda_{\delta_0}(j))\ge 1 \quad \text{for all} \quad  j \in J_{e},
  \eeq 
 we have  
 \begin{equation}\label{lb1}
H_{X,(Y, t_{Y}),\Lambda} \ge 2 C_{u }  {\delta_{0}}^{-2(d+1)}\quad  \text{on} \quad  \L^{2}(\Lambda).
\end{equation}
Setting   $E_{0}=  C_{u}{\delta_0^{-2(d+1)}}$, it follows that
for all $E \in [0, E_{0}]$ we get
\begin{align}\label{wegL0}
\| R_{X,(Y,t_{Y}),\Lambda}(E) \|& \le  E_{0}^{-1}
\intertext{and}   \label{goodL01}
\| \chi_y  R_{X,(Y,t_{Y}),\Lambda}(E)  \chi_{y^{\prime}} \|& \le 
2 E_{0}^{-1} \e^{- \sqrt{E_{0}}|y-y^{\prime}|},\;  \text{for $y,y^{\prime} \in \Lambda$ with $\abs{y-y^{\pr}} \ge 4 \sqrt{d}$}.
 \end{align}
\end{lemma}

\begin{proof} 
 Given  configurations $X$ and $Y$ such that $X\cap Y=\emptyset$ and $X$ satisfies \eqref{XinJe},
 we pick $\zeta_{j} \in X_{\Lambda_{\delta_{0}}(j)} $ for each $j \in J_{e}$, and set  $X_{1}: = \{\zeta_{j}, \ j \in J_{e} \}$,
$X_{2}= (X \setminus X_{1})\sqcup Y$. We claim that for all $t_{X_{2}}$ we have 
\begin{equation}\label{lb19}
H_{X_{1},(X_{2}, t_{X_{2}}),\Lambda} \ge H_{ X_{1},\Lambda}\ge 2 C_{u }  {\delta_{0}}^{-2(d+1)}\quad  \text{on} \  \L^{2}(\Lambda),
\end{equation}
where $C_{u}>0$.  Although the first inequality is obvious, the second is not, since 
\begin{equation}
|\{V_{X_{1}}\not=0\}|  \le  L^{d}\delta_{+}^{d}\delta_{0}^{d} < L^{d} \quad \text{if $\delta_{0} > \delta_{+}$}.
\end{equation}
To overcome this lack of a strictly positive  bound from below for $V_{X_{1}}$ on $\Lambda$,  we use the averaging procedure introduced in \cite{BK}.
  Requiring $\delta_{0} > \delta_{-}$, we have
\begin{equation}\label{average}
\overline{V}_{X_{1}}(y) := \frac1{(6{\delta_{0}})^d} \int_{\Lambda_{6 {\delta_{0}}}(0)}  \mathrm{d} a \, V_{X_{1}} (y-a)  \ge  c_{u} \,{{\delta_{0}}^{-d}} \chi_{\Lambda}(y) \quad \text{with $c_{u} >0$},
\end{equation}
by the definition of $X_{1}$ plus the lower bound in \eqref{u},
and hence
\begin{equation}\label{lb2}
\overline{ H}_{ X_{1},\Lambda}:= -\Delta_{\Lambda } +\chi_{\Lambda}\overline{V}_{X_{1}} \ge  c_{u}{{\delta_{0}}^{-d}} \quad \text{on} \quad \L^{2}(\Lambda). 
\end{equation}
Thus, if $\vphi \in C_{c}^{\infty}({\Lambda})$ with $\norm{\vphi}=1$, we have
\begin{align}\notag
& \scal{\vphi,H_{ X_{1},\Lambda}\vphi }_{\Lambda}= \scal{\vphi,\overline{H}_{ X_{1},\Lambda}\vphi }_{\Lambda}
+ \scal{\vphi,\left({V}_{X_{1}}- \overline{V}_{X_{1}}\right) \vphi }_{\Lambda}\\
&\quad \ge c_{u}{{\delta_{0}}^{-d}} +  \scal{\vphi,\left({V}_{X_{1}}- \overline{V}_{X_{1}}\right) \vphi }_{\R^{d}}\\
& \quad \ge c_{u}{{\delta_{0}}^{-d}} + \scal{\vphi,{V}_{X_{1}}\vphi }_{\R^{d}}- \frac1{(6{\delta_{0}})^d} \int_{\Lambda_{6 {\delta_{0}}}(0)}  \mathrm{d} a \,\scal{\vphi(\cdot +a),{V}_{X_{1}}\vphi(\cdot +a) }    \notag\\
& \quad \ge c_{u}{{\delta_{0}}^{-d}} -\frac1{(6{\delta_{0}})^d} \int_{\Lambda_{6 {\delta_{0}}}(0)}  \mathrm{d} a \, \left|\scal{\vphi,{V}_{X_{1}}\vphi }- \scal{\vphi(\cdot +a),{V}_{X_{1}}\vphi(\cdot +a) }\right|     \notag\\
& \quad \ge c_{u}{{\delta_{0}}^{-d}} - c^{\prime}_{u} {\delta_{0}}\norm{\nabla_{\Lambda}\vphi}_{\Lambda}
\ge c_{u}{{\delta_{0}}^{-d}} - c^{\prime}_{u}{\delta_{0}} 
 \scal{\vphi,H_{ X_{1},\Lambda}\vphi }_{\Lambda}^{\frac 1 2},\notag
\end{align}
where we used
\begin{equation}
\norm{\vphi(\cdot +a)-\vphi}_{\R^{d}}= \norm{(\e^{ a\cdot \nabla} -1)\vphi}_{\R^{d}}\le \abs{a} \norm{\nabla\vphi}_{\R^{d}}= \abs{a} \norm{\nabla_{\Lambda}\vphi}_{\Lambda}.
\end{equation}
It follows that there is $ \tilde{\delta}_{u}\ge \delta_{-}$, such that  for $\delta_{0}>\tilde{\delta}_{u}$ we have 
\begin{equation}
 \scal{\vphi,H_{ X_{1},\Lambda}\vphi }_{\Lambda} \ge c^{\prime\prime}_{u}\,{{\delta_{0}}^{-2(d+1)}},
\end{equation}
and hence we get    \eqref{lb19}, which implies \eqref{lb1}.

If we now set $E_{0}=  C_{u}{\delta_0^{-2(d+1)}}$,
then for all   $E \in [0, E_{0}]$ we get \eqref{wegL0} 
  immediately from \eqref{lb1},  and  \eqref{goodL01}   follows from \eqref{lb1} by the Combes-Thomas estimate (we use the precise estimate in \cite[Eq. (19)]{GK2}).
  \end{proof}

\begin{proof}[Proof of Proposition~\ref{proppoisson}] 
Given $\vrho>0$,  $p>0$, let $C_{u}$ and $\tilde{\delta}_{u}$ be the constant from Lemma~\ref{lempoisson}, and for scales $L>1$ let $\delta_L, E_{L}$, and $m_{L}$ be as in \eqref{EmL}. Given a box  $\Lambda=\Lambda_{L}(x)$, let  $J,J_{e} $  be as in Lemma~\ref{lempoisson} with $\delta_{0}=\delta_{L}$,  and set $\Lambda^{(e)}=\bigcup_{j \in J_{e}} \Lambda_{\delta_{L}}(j)$. We require 
\beq \label{vrhoL1}
 \vrho \le  (p+d+1)\tilde{\delta}_{u}^{-d}  \log L , \quad \text{which implies}  \quad 
\delta_{L}\ge 1 + \tilde{\delta}_{u},  \quad \text{and} \quad L> \delta_{L}+ \delta_{+}.
\eeq

We let $\widehat{\cJ}_{\Lambda} $ denote the collection of all   $(B,B^{\pr},S)\in{\cJ}_{\Lambda} $ such that
\begin{gather}
B\sqcup B^{\pr}\sqcup S \in {\cJ}_{\Lambda}, \quad B\sqcup B^{\pr}\subset \Lambda^{(e)}, \quad S \cap \Lambda^{(e)}=\emptyset ;\\  \label{NBj}
 N_{B}(\Lambda_{{\delta_{L}}}(j))\ge 1 \quad \text{for all} \quad  j \in J_{e}; \\
  N_{S}(\Lambda_{{\delta_{L}}}(j))\ge 1 \quad \text{for all} \quad  j \in J \setminus J_{e} . \label{NSj}
\end{gather}
If $(B,B^{\pr},S)\in \widehat{\cJ}_{\Lambda} $,  it is a consequence of \eqref{NSj} that the density condition \eqref{densitycond} holds for $S$ in $\Lambda$ if 
\beq \label{vrhoL2}
\vrho \ge c_{p,d} L^{-(0+)}, \quad \text{where} \quad c_{p,d}> 0,
\eeq
and then it follows from \eqref{NBj} and  Lemma~\ref{lempoisson}  that  $ \cC_{\Lambda,B,B^{\pr},S}$ is a $(\Lambda,E,m_{L})$-adapted bevent for all $E \in [0,E_{L}]$  if we also have
 \beq \label{vrhoL3}
\vrho \ge c_{p,d,u} L^{-\frac d {d +3}}, \quad \text{where} \quad c_{p,d,u}> 0.
 \eeq
Moreover, if $(B_{i},B_{i}^{\pr},S_{i})\in \widehat{\cJ}_{\Lambda}$, $i=1,2$, and
$(B_{1},B_{1}^{\pr},S_{1})\not= (B_{2},B_{2}^{\pr},S_{2})$, then  $ \cC_{\Lambda,B_{1},B_{1}^{\pr},S_{1}}\cap  \cC_{\Lambda,B_{2},B_{2}^{\pr},S_{2}}=\emptyset$.  We conclude that
\beq
\Omega_{\Lambda}=\bigsqcup_{(B,B^{\pr},S)\in \widehat{\cJ}_{\Lambda}} \cC_{\Lambda,B,B^{\pr},S}
\eeq
is a $(\Lambda,E,m_{L})$-localizing event   $E \in [0,E_{L}]$ if \eqref{vrhoL1}, \eqref{vrhoL2} and \eqref{vrhoL3} are satisfied, which can be assured by requiring that
 $L >\overline{L}_{1}(d,u,\vrho,p) $.

To establish \eqref{PEL0}, let $\delta_{L}^{\pr}:=\delta_{L}-1=( (p+d+1){ \vrho}^{-1} \log L)^{\frac1d}$, and consider the event
\beq
\Omega_{\Lambda}^{(\ddagger)}:=\{ N_{\X}(\Lambda_{\delta_{L}^{\pr}}(j))\ge 1 \quad \text{for all} \quad  j \in J \}.
\eeq
Clearly
  \begin{equation}\label{PE1}
\P \{\Omega_{\Lambda}^{(\ddagger)} \}\ge 1 - \left(\tfrac {L}{ {\delta_{L}}}  \right)^{d}\e^{-\vrho( \delta_{L}^{\pr})^{d}}\ge 1 -  L^{- p-1}.
\end{equation}
Since $\delta_{L}-\delta_{L}^{\pr}=1 \ge \eta_{L}$, we must have
\beq
\Omega_{\Lambda}^{(\ddagger)} \cap \Omega_{\Lambda}^{(0)}\subset \Omega_{\Lambda},
\eeq
and hence \eqref{PEL0} follows from \eqref{PE1} and \eqref{Omega0} for $L >\overline{L}_{0}(d,u,\vrho,p) $ satisfying \eqref{vrhoL}.
\end{proof}

\subsection{Fixed interval at the bottom of the spectrum and high disorder}

Proposition~\ref{proppoisson} can also be formulated  for  a fixed interval at the bottom of the spectrum and high disorder.  

\begin{proposition}\label{proppoissonbis}  
 Let $H_{\X}$ be a Poisson Hamiltonian on  $\L^{2}(\R^{d})$ with density $\vrho >0$, and  fix $p>0$.   
 Given $E_{0}>0$,  there   exist   a constant  $C_{d,u,p,E_{0}}>0$
and a scale  $\overline{L}_{0}=\overline{L}_{0}(d,u,E_0,p) < \infty$, such that if  $L\ge\overline{ L}_{0}$  and  $\vrho \ge C_{d,u,p,E_{0}} \log L$ satisfy  \eqref{vrhoL},
 setting  $m=\frac 1 2 \sqrt{E_{0}}$,
 the  scale $L$  is $(E,m)$-localizing for all energies $E \in [0,E_{0}]$.
 \end{proposition}

\begin{proof} Given $E_{0}>0$ and $ p>0$, let $K_0=\min \{k \in \N; \, k\ge 2 u_{-}^{{-1}} E_{0}\}$,  $\Lambda=\Lambda_{L}(x)$, fix ${\delta_{0}}= \frac 1 6 {\delta_{-}} $, and let  $ J,J_{e},\Lambda^{(e)} $  be as in Proposition~\ref{proppoisson} (with $\delta_{0}$ instead of $\delta_{L}$).
Given $ X ,Y\in \cP_{0}(\R^{d})$ and  $t_{Y}\in [0,1]^{Y}$, such that $X \cap Y=\emptyset$ and
  \beq  \label{XinJe2}
   N_{X}(\Lambda_{\delta_0}(j))\ge K_{0} \quad \text{for all} \quad  j \in J_{e},
  \eeq 
 we have  
\begin{equation}\label{lb198}
H_{X,(Y, t_{Y}),\Lambda} \ge 2 E_{0}\quad  \text{on} \quad  \L^{2}(\Lambda),
\end{equation}
and \eqref{wegL0}  and  \eqref{goodL01} follows as in  Lemma~\ref{lempoisson}.

To prove \eqref{lb198},    fix $X_{1} \subset X$ such that  has exactly $K_{0}$ points in each box $\Lambda_{\delta_0}(j)$ for all $j \in J_{e}$ and none outside these boxes, that is, 
\beq \begin{split}
 N_{X_{1}}(\Lambda_{\delta_0}(j))= K_{0} \; \text{for all} \;  j \in J_{e} \quad 
 \text{and} \quad  N_{X_{1}}(\R^{d}\setminus\Lambda^{(e)}  )=0.
\end{split} \eeq
  By our choice of ${\delta_{0}}$  and \eqref{u} we get
  \begin{equation}\label{newbound}
{V}_{X_{1}}(y)  \ge  K_{0} u_{-} \chi_{\Lambda}(y)\ge 2 E_{0}\chi_{\Lambda}(y),
\end{equation}
and hence, setting  $X_{2}=X\setminus X_{1}$,  for all $ t_{X_{2}}\in [0,1]^{X_{2}}$  we have
\begin{equation}\label{lb12}
H_{X_{1},(X_{2}, t_{X_{2}}),\Lambda} \ge H_{X_{1},\Lambda}\ge 2 E_{0},
\end{equation}
and \eqref{lb198} follows.

We now modify the argument in the proof  of  Proposition~\ref{proppoisson}.
Let $\widehat{\cJ}_{\Lambda} $ denote the collection of all   $(B,B^{\pr},S)\in{\cJ}_{\Lambda} $ such that
\begin{gather}
B\sqcup B^{\pr}\sqcup S \in {\cJ}_{\Lambda}, \quad B\sqcup B^{\pr}\subset \Lambda^{(e)}, \quad S \cap \Lambda^{(e)}=\emptyset ;\\  \label{NBj76}
 N_{B}(\Lambda_{\delta_0}(j))\ge K_{0} \quad \text{for all} \quad  j \in J_{e}; \\
  N_{S}(\Lambda_{\delta_0}(j))\ge 1 \quad \text{for all} \quad  j \in J \setminus J_{e} . \label{NSj98}
\end{gather}
If $(B,B^{\pr},S)\in \widehat{\cJ}_{\Lambda} $, the density condition \eqref{densitycond} for $S$ in $\Lambda$  follows from \eqref{NSj98}, and it follows from \eqref{NBj76} and  \eqref{lb198}  that  $ \cC_{\Lambda,B,B^{\pr},S}$ is a $(\Lambda,E,m)$-adapted bevent with  $m=\frac 1 2 \sqrt{E_{0}}$  for all $E \in [0,E_{0}]$ if $L\ge\overline{ L}_{1}(u,E_{0})$. 
  We conclude that
\beq
\Omega_{\Lambda}=\bigsqcup_{(B,B^{\pr},S)\in \widehat{\cJ}_{\Lambda}} \cC_{\Lambda,B,B^{\pr},S}
\eeq
is a $(\Lambda,E,m)$-localizing event for all  $E \in [0,E_{0}]$. 

To establish \eqref{PEL0}, let $\delta_{1}:=\frac 1 2\delta_{0}$ and consider the event
\beq
\Omega_{\Lambda}^{(\ddagger)}:=\{ N_{\X}(\Lambda_{\delta_1}(j))\ge K_{0} \quad \text{for all} \quad  j \in J \}.
\eeq
  We have, using \eqref{Poissonk012},
 \begin{equation}\label{PElarge}
\P \{\Omega_{\Lambda}^{(\ddagger)} \}\ge 1 -{\left(\tfrac {L}{ {\delta_{0}}}  \right)^{d}}C_{K_{0}} \mathrm{e}^{-\frac 1 2 \varrho\delta_{1}^{d}}=  
1 -C_{u,E_{0},d}L^{d} \mathrm{e}^{- c_{u,d} \varrho} \ge 1 -  L^{- p-1}
\end{equation}
for $\vrho \ge C_{d,u,p,E_{0}} \log L$ if $L\ge\overline{ L}_{2}(u,E_{0},d,p)$

Since $\delta_{0}-\delta_{1}=\frac 1 {12} \delta_{-} \ge \eta_{L}$ for $L\ge\overline{ L}_{3}(u)$, for $L\ge\overline{ L}_{4}(u,E_{0},d,p)$ we must have
\beq
\Omega_{\Lambda}^{(\ddagger)} \cap \Omega_{\Lambda}^{(0)}\subset \Omega_{\Lambda},
\eeq
and hence \eqref{PEL0} follows from \eqref{PElarge} and \eqref{Omega0} for $L >\overline{L}_{0}(d,u,E_{0},p) $ with $\vrho \ge C_{d,u,p,E_{0}} \log L$.
\end{proof}

\section{The multiscale analysis with  a Wegner estimate}
\label{sectMSA}

We can now state our version of \cite[Proposition~A$^{\!\prime}$]{BK} for Poisson Hamiltonians.

\begin{proposition}  \label{propA} Let $H_{\X}$ be a Poisson Hamiltonian on  $\L^{2}(\R^{d})$ with density $\vrho >0$.  Fix an energy $E_{0} >0$.   Pick      $p=
\frac 3 8 d -$,  $  \rho_{1}=\frac 3 4 -$ and
$ \rho_{2}=0+$,  more precisely, pick $p, \rho_{1}, \rho_{2}$ such that 
\beq  \label{rhos}
\tfrac 8 {11} < \tfrac d {d +p} <  \rho_{1}< \tfrac 3 4, \quad
 \text{ $ \rho_{2}=\rho_{1}^{n_{1}}$ with $n_{1}\in  \N$ and $ p < d (\tfrac {\rho_{1}} 2 -\rho_{2})$}.
\eeq
  Let $ E \in [0,E_{0}]$, and suppose   $L$ is $(E,m_{0})$-localizing for all 
  $L \in [L_{0}^{\rho_{1}\rho_{2}}, L_{0}^{\rho_{1}}]$, where
 \beq \label{m0}
 m_{0}\ge  L_{0}^{-\tau_{0}} \quad \text{with} \quad \tau_{0}=0+ < \rho_{2},
 \eeq
the condition  \eqref{vrhoL} is satisfied at scale $L_{0}^{\rho_{1}\rho_{2}}$, 
 and the scale $L_{0}$   is  also sufficiently large 
(depending on $d, E_{0},p, \rho_{1},\rho_{2},\tau_{0}$) .  Then $L$ is $(E,\frac  {m_{0}} 2)$-localizing for all $L \ge L_{0}$ (actually, for all $L \ge L_{0}^{\rho_{1}\rho_{2}}$).
\end{proposition}

The proof will require several lemmas and definitions.

\begin{lemma} \label{lemlocLell} Fix $p^{\pr}=p-$ and let $\Lambda_{\ell}\subset \Lambda=\Lambda_{L}$ with $\ell \ll L$.  If the scale $\ell$ is   $(E,m)$-localizing, then   there exists 
a $(\Lambda,\Lambda_{\ell}, E,m)$-localized event $\Omega_{\Lambda}^{\Lambda_{\ell}}$, i.e., 
\beq \label{adaptedbox}
\Omega_{\Lambda}^{\Lambda_{\ell}}= \bigsqcup_{i=1}^{I_{L,\ell}} \cC_{\Lambda,B_{i},B_{i}^{\pr},S_{i}}^{\Lambda_{\ell}}
\eeq
for some 
  disjoint $(\Lambda,\Lambda_{\ell},E,m)$-adapted bevents
$\{\cC_{\Lambda,B_{i},B_{i}^{\pr},S_{i}}^{\Lambda_{\ell}}\}_{i=1,2,\ldots,I_{L,\ell}}$,
such that
\beq \label{adaptedbox22}
\P\{\Omega_{\Lambda}^{\Lambda_{\ell}}\} > 1- \ell^{-p^{\pr}} .
\eeq
\end{lemma}

\begin{proof}
Given  disjoint $\Lambda_{\ell}$-bevents, the  corresponding 
 $(\Lambda,\Lambda_{\ell})$-bevents in \eqref{cylinderLL1}  are also disjoint events.  
 Since the scale $\ell$ is   $(E,m)$-localizing, there is a $(\Lambda_{\ell},E,m)$-localized event $\Omega_{\Lambda_{\ell}}$ satisfying \eqref{PEL0}.  From Lemma~\ref{lemredraw} we get
 \beq
 \Omega_{\Lambda_{\ell}} \cap \Omega_{\Lambda}^{(0)}\subset\Omega_{\Lambda}^{\Lambda_{\ell}},
 \eeq
 where $\Omega_{\Lambda}^{\Lambda_{\ell}}$ is as in \eqref{adaptedbox}.
 The estimate \eqref{adaptedbox22} then follows from  \eqref{PEL0} and  \eqref{Omega0}.
\end{proof}

\begin{definition}
Given scales $\ell\le  L$,  a standard $\ell$-covering of a box $\Lambda_{L}(x)$ is
a collection of boxes $\Lambda_{\ell}$ of the form
\begin{align}\label{standardcover}
\cG_{\Lambda_{L}(x)}^{(\ell)}= \{ \Lambda_{\ell}(r)\}_{r \in \G_{\Lambda_{L}(x)}^{(\ell)}},
\end{align}
where
\beq  \label{bbG}
\G_{\Lambda_{L}(x)}^{(\ell)}:= \{ x + \alpha\ell  \Z^{d}\}\cap \Lambda_{L}(x)\quad 
\text{with}  \quad \alpha \in ]\tfrac {3} {5},\tfrac {4} {5}]   \cap \{\tfrac {L-\ell}{2 \ell n}; \, n \in \N \}.
\eeq
\end{definition}

\begin{lemma}  If $\ell \ll L$ there is always a   standard $\ell$-covering
$\cG_{\Lambda_{L}(x)}^{(\ell)}$ of a box $\Lambda_{L}(x)$, and    we have
 \begin{align}\label{nestingproperty}
&\Lambda_{L}(x) =\bigcup_{r \in\G_{\Lambda_{L}(x)}^{(\ell)} } \Lambda_{\ell}(r),\\ \label{bdrycover}
&\text{for each $y \in \Lambda_{L}(x)$ there is $r \in \G_{\Lambda_{L}(x)}^{(\ell)}$ with $\Lambda_{\frac{2\ell} 5}(y)\cap \Lambda_{L}(x) \subset \Lambda_{\ell}(r)$},\\
\label{freeguarantee}
&\Lambda_{\frac{\ell}{5}}(r)\cap \Lambda_{\ell}(r^{\prime})=\emptyset
\quad \text{if} \quad r\ne r^{\prime},\\ \label{number}
& \#\G_{\Lambda_{L}(x)}^{(\ell)} \le   (\tfrac {5} {3}\tfrac{L} {\ell})^{d}\le   (\tfrac{2L} {\ell})^{d}.
\end{align}
Moreover we have  the following nesting property:  Given $y \in  x + \alpha \ell  \Z^{d}$ and $n \in \N$ such that $\Lambda_{(2  n \alpha  +1) \ell}(y)\subset \Lambda$, it follows that
\beq \label{nesting}
\Lambda_{(2  n \alpha  +1) \ell}(y)= \bigcup_{r \in  \{ x + \alpha \ell \Z^{d}\}\cap\Lambda_{(2n \alpha  +1) \ell}(y) } \Lambda_{\ell}(r),
\eeq
and  $ \{ \Lambda_{\ell}(r)\}_{r \in  \{ x + \alpha\ell  \Z^{d}\}\cap\Lambda_{(2n\alpha +1) \ell}(y)}$ is a standard $\ell$-covering of the box $\Lambda_{(2 n\alpha +1) \ell}(y)$.
\end{lemma}

\begin{proof}
The lemma can be easily checked using \eqref{bbG}. In particular, 
$\alpha>\frac 35$ ensures \eqref{bdrycover}, $\alpha\le\frac 45$ ensures \eqref{freeguarantee}, and the existence of $n\in \N$ such that
$2 n \alpha \ell = L - \ell $ ensures the nesting 
property \eqref{nestingproperty}.
\end{proof}

 In the following we  fix $ E \in [0,E_{0}]$, assume \eqref{rhos}, and
set  $\Lambda=\Lambda_{L}$,  $\ell_{1}=L^{\rho_{1}}$, and  $\ell_{2}=L^{\rho_{2}}$.  
We also assume the  induction hypotheses: for each box $\Lambda_{{\ell}}\subset \Lambda$ with
${\ell}\in [\ell_{2},\ell_{1}]$ 
there is  a $(\Lambda_{\ell},E,m_{0})$-localized event  $\Omega_{\Lambda_{{\ell}}}$ with
 \eqref{PEL0}, and hence it follows from Lemma~\ref{lemlocLell} that there is a 
  $(\Lambda,\Lambda_{\ell}, E,m_{0})$-localized event $\Omega_{\Lambda}^{\Lambda_{\ell}}$ with \eqref{adaptedbox22}, and we have \eqref{m0} with $m_{0}$ and $L$.  

 \begin{remark}
  The rate of decay $m$ in  \eqref{good}, which by hypothesis is  $m_{0}$ as in \eqref{m0} for all scales $L \in [L_{0}^{\rho_{1}\rho_{2}}, L_{0}^{\rho_{1}}]$,
 will vary along the multiscale analysis, i.e., the construction gives a rate of decay $m_{L} $ at scale $L$.   The control of this variation can be done as usual, as commented in \cite{BK} (but we need a condition like \eqref{m0}),  so we always have
 $m_{L}\ge\frac  {m_{0}} 2$ , e.g.,  \cite{vDK,FK,GK1,Kle}).  We will ignore this variation as in \cite{BK} and simply write  $m$ for $m_{L}$. We will omit $m$ from the notation in the rest of this section.  The exponent  $1-$ in \eqref{weg} does not vary.
  \end{remark}

We now define an event that incorporates  \cite[property $(\ast)$]{BK}.

\begin{definition}
 Given a box $\Lambda_{\ell_{1}}$, for each  $n=0,1,\ldots,n_{1}$ let $L_{n}=:\ell_{1}^{\rho_{1}^{n}}$  (note $L_{0}=\ell_{1} $, $L_{n_{1}}=\ell_{2}$),
and  let $\cR_{n}=\{\Lambda_{L_{n}}(r)\}_{r \in R_{n}}$ be a standard  $L_{n}$-covering  of  $\Lambda_{\ell_{1}}$ as in \eqref{standardcover}.   For a given number $K_{2}$, 
a configuration set $X$ is said to be   $(\Lambda_{\ell_{1}},E)$-notsobad   if   there is $\Upsilon_{B}=\cup_{r \in R^{\pr}_{n_{1}}} \Lambda_{3\ell_{2}}(r)$, where $R^{\pr}_{n_{1}} \subset  R_{n_{1}}$ with $\# R^{\pr}_{n_{1}} \le K_{2}$,  
 such that for all $ x \in   {\Lambda_{\ell_{1}}}\setminus \Upsilon_{B}$ there is an
 $(X,E)$-jgood box $\Lambda_{L_{n}}(r)$, with   $r\in R_{n}$ for some $n\in \{1,\ldots,n_{1}\}$ and $ \Lambda(x, \frac {2 L_{n}} 5)  \cap \Lambda_{\ell_{1}}\subset \Lambda_{L_{n}}(r)$. If $\Lambda_{\ell_{1}} \subset \Lambda$, a $(\Lambda,\Lambda_{\ell_{1}})$-bconfset $C_{\Lambda,B}^{\Lambda_{\ell_{1}}}$ or bevent $C_{\Lambda,B,B^{\pr}}^{\Lambda_{\ell_{1}}}$ is  $(\Lambda,\Lambda_{\ell_{1}},E)$-notsobad  if
 the configuration set $B$ is $(\Lambda_{\ell_{1}},E)$-notsobad.
\end{definition}

\begin{lemma} \label{lemast}  For sufficiently large $K_{2}$,
depending only on $d,p,\rho_{1}, n_{1}$, for all boxes   $\Lambda_{\ell_{1}}\subset \Lambda$, with $\ell_{1}$ large enough,  there exist  
  disjoint $(\Lambda,\Lambda_{\ell_{1}},E)$-notsobad bevents
$\{\cC_{\Lambda,B_{m},B_{m}^{\pr}}^{\Lambda_{\ell_{1}}}\}_{m=1,2,\ldots,M}$  such that 
\beq \label{notsobad}
\P\{\Omega_{\Lambda}^{\Lambda_{\ell_{1}},(*)}\} > 1- \ell_{1}^{-5d},
\qquad \text{with}
\qquad
\Omega_{\Lambda}^{\Lambda_{\ell_{1}},(*)}= \bigsqcup_{m=1}^{M} \cC_{\Lambda,B_{m},B_{m}^{\pr}}^{\Lambda_{\ell_{1}}} ,
\eeq
and hence
\beq \label{notsobad3}
\Omega_{\Lambda}^{\Lambda_{\ell_{1}},(*\setminus)}:= \Omega_{\Lambda}^{\Lambda_{\ell_{1}},(*)}\setminus \Omega_{\Lambda}^{\Lambda_{\ell_{1}}}= 
\bigsqcup_{q=1}^{Q} \cC_{\Lambda,F_{q},F_{q}^{\pr}}^{\Lambda_{\ell_{1}}},
\eeq
where $\{\cC_{\Lambda,F_{q},F_{q}^{\pr}}^{\Lambda_{\ell_{1}}}\}_{q=1,2,\ldots,Q}$ are disjoint  $(\Lambda,\Lambda_{\ell_{1}},E)$-notsobad bevents.
\end{lemma}

\begin{proof}  Given $\Lambda_{L_{n-1}}(r) \in \cR_{n-1}$, we set
\beq \begin{split}
\cR_{n}(r)& := \{\Lambda_{L_{n}}(s) \in \cR_{n}; \,\Lambda_{L_{n}}(s)\cap \Lambda_{L_{n-1}}(r)\not= \emptyset \} \quad \text{and} \quad\\
 R_{n}(r)&:=\{s\in \cR_{n}; \, \Lambda_{L_{n}}(s)\in \cR_{n}(r)\}.
\end{split}
\eeq
We  have  $\Lambda_{L_{n-1}}(r) \subset \bigcup_{s \in R_{n}(r)} \Lambda_{L_{n}}(s) $
and, similarly to \eqref{number}, $\#  R_{n}(r) \le  (\frac{3L_{n-1}} {L_{n}})^{d}$.  Fix  a number $K^{\pr}$, and define the event $\Omega_{\Lambda}^{\Lambda_{\ell_{1}},(*\pr)}$ as consisting of $\omega \in \Omega$ such that,
for all $n=1,\dots,n_{1}$ and all $r\in R_{n-1}$, we have
$\omega \in \Omega_{\Lambda}^{\Lambda_{L_{n}}(s)}$ for all $s \in R_{n}(r)$, with the possible exception of at most $K^{\pr}$ disjoint boxes $\Lambda_{L_{n}}(s)$ with  $s \in R_{n}(r)$.  The probability of its complementary event can be estimated from \eqref{adaptedbox22}  as in  \cite[Eq. (6.12)]{BK}:
\begin{align} \label{probast}
\P\left\{\Omega \setminus \Omega_{\Lambda_{\ell_{1}}}^{(*\pr)}\right \}
&\le \sum_{n=1}^{n_{1}}    (\tfrac{2\ell_{1}} {L_{n-1}})^{d}   (\tfrac{3L_{n-1}} {L_{n}})^{K^{\pr} d}   L_{n}^{-K^{\pr}p^{\pr}}\\ \notag
& \le 2^{d} 3^{K^{\pr}d} n_{1 }
 \ell_{1}^{-\rho_{1}^{n_1 -1}(K^{\pr}(\rho_{1}(p^{\pr}+d)-d)+d)+ d}
 \le \ell_{1}^{-6d},
\end{align}
which holds for all large $\ell_{1}$ after choosing $K^{\pr}$  sufficiently large using \eqref{rhos}.

Given  $\omega \in\Omega_{\Lambda}^{\Lambda_{\ell_{1}},(*\pr)}$, then for each  $n=1,\dots,n_{1}$ and  $r\in R_{n-1}$ we can find $s_{1},s_{2},\ldots,s_{K^{\pr\pr}}\in R_{n}(r)$, with $K^{\pr\pr}\le K^{\pr}-1$, such that $\omega \in   \Omega_{\Lambda}^{\Lambda_{L_{n}}(s)}$  if $s \in R_{n(r)}$
and $s \notin \bigcup_{j=1}^{K^{\pr\pr}} {\Lambda_{3L_{n}}(s_{j})}$.  
(Here we need boxes of side $3L_{n}$ because we only ruled out the existence of $K^{\pr}$ \emph{disjoint} boxes of side $L_{n}$.)  Since each box $ {\Lambda_{3L_{n}}(s_{j})}$ is contained in the union of  at most $C^{\pr\pr} $ boxes in 
$\cR_{n}$, we conclude that for each $\omega \in\Omega_{\Lambda}^{\Lambda_{\ell_{1}},(*\pr)}$ there are  $t_{1},t_{2},\ldots,t_{K^{\pr\pr\pr}} \in R_{n_{1}}$, with   $K^{\pr\pr\pr}\le K_{2}=(C^{\pr\pr} (K^{\pr}-1))^{n_{1}} $, such that , setting $\Upsilon= \bigcup_{t_{j}=1}^{K^{\pr\pr\pr}} \Lambda_{3\ell_{2} }(t_{j})$,  for all $ x \in   {\Lambda_{\ell_{1}}}\setminus \Upsilon$ we have $\omega \in   \Omega_{\Lambda}^{\Lambda_{L_{n}}(s)}$  for some  $n=1, 2,\ldots,n_{1}$ and  $s \in R_{n}$, with  and $ \Lambda(x, \frac {2 L_{n}} 5)  \cap \Lambda_{\ell_{1}}\subset \Lambda_{L_{n}}(s)$.

Recalling \eqref{unionOmega0}, we have 
\beq
\Omega_{\Lambda}^{\Lambda_{\ell_{1}},(*\pr)} \cap \Omega_{\Lambda}^{(0)}
\subset \Omega_{\Lambda}^{\Lambda_{\ell_{1}},(*)}:=\bigsqcup_{\{(F,F^{\pr}); \, F\sqcup F^{\pr}\in\cJ_{\Lambda}^{\Lambda_{\ell_{1}}},\,  \cC_{\Lambda,F,F^{\pr}}^{\Lambda_{\ell_{1}}} \cap\Omega_{\Lambda}^{\Lambda_{\ell_{1}},(*\pr)}  \not= \emptyset\}}
 \cC_{\Lambda,F,F^{\pr}}^{\Lambda_{\ell_{1}}}.
\eeq
It follows from Lemma~\ref{lemqgood} that each  $ \cC_{\Lambda,F,F^{\pr}}$ in the disjoint union must be a $(\Lambda,\Lambda_{\ell_{1}},E)$-notsobad bevent. Thus \eqref{notsobad} follows from \eqref{probast} and \eqref{Omega0}. We obtain 
 \eqref{notsobad3} from \eqref{notsobad} and \eqref{goodcylinders}.
\end{proof}

\begin{definition}
Let 
 $\cR=\{\Lambda_{\ell_{1}}(r)\}_{r \in R}$ be a standard $\ell_{1}$-covering of  $\Lambda$ and fix   $K_{1} \in \N$.  A 
 $\Lambda$-bevent $\cC_{\Lambda,B,B^{\pr},S}$,  is called     $(\Lambda,E)$-prepared if 
 $S$ satisfies the density condition
\begin{equation}  \label{densitycond2}
\# (S\cap {\widehat{\Lambda}_{\ell}} )\ge \ell^{d-} , \quad \text{for all
boxes}\quad   \Lambda_{\ell}\subset \Lambda\quad \text{with} \quad \ell_{1}\ll \ell \le L,
\end{equation}
and there is $R^{\prime}\subset R$ with $ \#(R \setminus R^{\prime})\le K_{1} $,  such that  if $r \in R^{\prime}$ then  
 $\cC_{\Lambda,B,B^{\pr},S}^{\Lambda_{\ell_{1}}(r)}$ is  a $(\Lambda,\Lambda_{\ell_{1}}(r),E)$-adapted  bevent, and if $r \in R\setminus R^{\prime}$ then
$S \cap \Lambda_{\ell_{1}}(r)=\emptyset $ and $\cC_{\Lambda,B,B^{\pr}}^{\Lambda_{\ell_{1}}(r)}$ is  a $(\Lambda,\Lambda_{\ell_{1}}(r),E)$-notsobad bevent.
\end{definition}

\begin{lemma} \label{lemstructure}Let 
 $\cR=\{\Lambda_{\ell_{1}}(r)\}_{r \in R}$ be a standard $\ell_{1}$-covering of  $\Lambda$.
For sufficiently large $K_{1}$,
depending only on $d,p,\rho_{1}, n_{1}$, if  $L$ is taken large enough,   there exist  
  disjoint $(\Lambda,E)$-prepared  bevents
$\{\cC_{\Lambda,B_{m},B_{m}^{\pr},S_{m}}\}_{m=1,2,\ldots,M_{\Lambda}}$,  such that 
\beq \label{prepared}
\P\{\Omega_{\Lambda}^{(1)}\} > 1-2L^{-2d},
\qquad \text{with}
\qquad
\Omega_{\Lambda}^{(1)}= \bigsqcup_{m=1}^{M_{\Lambda}} \cC_{\Lambda,B_{m},B_{m}^{\pr},S_{m}}.
\eeq
\end{lemma}

\begin{proof}

Fix  $K_{1}$,  recall \eqref{adaptedbox}  and \eqref{notsobad3},  and define the event $ \Omega_{\Lambda}^{(1)}$ by the disjoint union
 \beq  \begin{split}
  \Omega_{\Lambda}^{(1)}& := \bigsqcup_{\substack{ R^{\prime}\subset R\\
  \#(R \setminus R^{\prime})\le K_{1}}}  \Omega_{\Lambda}^{(1)}(R^{\prime}), \quad\text{where} \\
 \Omega_{\Lambda}^{(1)}(R^{\prime})&=    \left\{ \bigcap_{r \in R^{\prime}} \Omega_{\Lambda} ^{\Lambda_{\ell_{1}}(r)} \right\} \bigcap \left\{ \bigcap_{r \in R\setminus R^{\prime}} \Omega_{\Lambda}^{\Lambda_{\ell_{1}}(r),(*\setminus)}  \right\} .
 \end{split} \eeq 
Using the probability estimates in \eqref{adaptedbox}  and \eqref{notsobad},  and taking $K_{1}$ sufficiently large (independently of the scale), we get
\beq  \label{Ld}
 \P\{\Omega_{\Lambda}^{(1)}\} > 1-2 L^{-2d}.
\eeq
This can be seen as follows.  First, from  \eqref{notsobad} and   \eqref{notsobad3}  we have 
\beq
\P\left\{ \Omega_{\Lambda} ^{\Lambda_{\ell_{1}}(r)} \cup  \Omega_{\Lambda}^{\Lambda_{\ell_{1}}(r),(*\setminus)}\right \}\ge \P\left\{  \Omega_{\Lambda}^{\Lambda_{\ell_{1}}(r),(*)}\right \} > 1- L^{-5 \rho_{1 }d},
\eeq
and hence 
\beq  \begin{split} \label{Ld1}
&\P\left\{ \bigcap_{r \in R}\left\{ \Omega_{\Lambda} ^{\Lambda_{\ell_{1}}(r)} \cup  \Omega_{\Lambda}^{\Lambda_{\ell_{1}}(r),(*\setminus)}\right\}\right\}> 1 -  \left( \frac {2L}{\ell_{1} }\right)^{d} L^{-5 \rho_{1 }d}\\
& \qquad  \qquad  \qquad \qquad  \ge 1 - 2^{d} L^{-(6 \rho_{1 }-1)d}>
1 - L^{-2d},
\end{split}
\eeq
for large $L$, using also \eqref{rhos}.  On the other hand, letting $K_{1}=C^{\pr}(K^{\pr}-1)$, it follows from \eqref{adaptedbox} and \eqref{rhos} that
\beq \begin{split} \label{Ld2}
&\P\left\{\text{there are $K^{\pr}$  disjoint boxes $\Lambda_{\ell_{1}}(r) \in \cR$
with $\omega \notin\Omega_{\Lambda} ^{\Lambda_{\ell_{1}}(r)}$}\right\}\\
& \qquad \qquad   \le   (\tfrac{2L} {\ell_{1}})^{d K^{\pr}}\ell_{1}^{-p^{\pr}K^{\pr}}  \le  2 ^{dK^{\pr}}
L^{-K^{\pr}(\rho_{1}(p^{\pr}+d) -d)}\le L^{-2d}
\end{split} \eeq
if $K_{1} > \tfrac {2dC^{\pr}} {\rho_{1}(p^{\pr}+d) -d}$ and $L$ is large enough.  Here $C^{\pr}$ is chosen such that the complementary has at most $K_{1}$ (not necessarily disjoint)   boxes $\Lambda_{\ell_{1}}(r) \in \cR$
with $\omega \notin\Omega_{\Lambda} ^{\Lambda_{\ell_{1}}(r)}$.
The estimate \eqref{Ld} follows from  \eqref{Ld1} and  \eqref{Ld2}.

Moreover, it follows from  \eqref{adaptedbox}  and \eqref{notsobad3}  that each   $\Omega_{\Lambda}^{(1)}(R^{\prime})$ is a disjoint union of (non-empty) events of the form
\beq \label{cDR}
\cD_{R^{\prime}}= \left\{ \bigcap_{r \in R^{\prime}}\cC_{\Lambda,B_{r},B_{r}^{\pr},S_{r}}^{\Lambda_{\ell_{1}}(r)} \right\} \bigcap \left\{ \bigcap_{r \in R\setminus R^{\prime}}\cC_{\Lambda,F_{r},F_{r}^{\pr} }^{\Lambda_{\ell_{1}}(r)}\right\},
\eeq
where 
 $\cC_{\Lambda,B_{r},B_{r}^{\pr},S_{r}}^{\Lambda_{\ell_{1}}(r)}$ is  a $(\Lambda,\Lambda_{\ell_{1}}(r),E)$-adapted  bevent for each $r \in R^{\prime}$, and
 $\cC_{\Lambda,F_{r},F_{r}^{\pr}}^{\Lambda_{\ell_{1}}(r)}$ is a  $(\Lambda,\Lambda_{\ell_{1}},E)$-notsobad bevent for each $r \in R\setminus R^{\prime}$.

It remains to show that $\cD_{R^{\prime}}$ can be written as a disjoint union 
of $(\Lambda,E)$-prepared  bevents. To do so let, as in \cite{BK}, let
\beq \label{SR'}
S_{R^{\prime}}:=\{ s \in \J_{\Lambda}; \, s \in \Lambda_{\ell_{1}}(r)\Rightarrow r \in R^{\prime} \ \text{and} \ s \in S_{r}  \}.
\eeq
Since \eqref{freeguarantee} yields
\beq
\bigcup_{r \in R^{\prime}} S_{r} \cap \Lambda_{\frac{\ell_{1}}{5}}(r)\subset S_{R^{\prime}},
\eeq
and $\#(R\setminus R^{\prime}) \le K_{1}$, it follows as in \cite[Eq.~(6.18)]{BK} that $S_{R^{\prime}}$ satisfies the density condition \eqref{densitycond2} in
$\Lambda$.  It follows from \eqref{cylinderexp2} and \eqref{SR'} that  we can rewrite the event   $\cD_{R^{\prime}}$ in \eqref{cDR}  as a disjoint union
\beq
\cD_{R^{\prime}}= \bigsqcup_{j \in J}   \cC_{\Lambda,A_{j},A_{j}^{\pr},S_{R^{\prime}}},
\eeq
where  $\{ \cC_{\Lambda,A_{j},A_{j}^{\pr},S_{R^{\prime}}}\}_{j \in J}$ are    $(\Lambda,E)$-prepared  bevents.
\end{proof}

We can now prove a Wegner estimate at scale $L$ using \cite[Lemma~5.1$^{\pr}$]{BK}.

\begin{lemma}\label{lemwegner}  Let   $\cC_{\Lambda,B,B^{\pr},S}$ be  a  $(\Lambda,E)$-prepared bevent, and consider a box $\Lambda_{L_{0}}\subset \Lambda $
with   $L_{0}=(2 n \alpha +1)\ell_{1}$ for some $n \in N$,   $\ell_{1} \ll L_{0}\le L$, such that $\Lambda_{L_{0}} $ is constructed as in \eqref{nesting} from  a standard $\ell_{1}$-covering of  $\Lambda$.  
Then, for sufficiently large  
$L$   there  exist disjoint subsets $\{S_{i}\}_{i=1,2,\dots,I}$ of $S_{0}:=S \cap \Lambda_{0}$, such that 
\beq  \label{wegner1}
\norm{R_{B\sqcup S_{i},\Lambda_{L_{0}}}(E)}< 
\e^{C_{1}L^{\frac 4 3 \rho_{1}}\log L  },\quad \text{for all} \quad  i=1,2,\dots,I,
\eeq
and we  have the  conditional probability estimate
\beq \begin{split} \label{wegner2}
& \P\{\left. \Omega_{\Lambda,B,B^{\pr},S}^{\Lambda_{0}}\right|\cC_{\Lambda,B,B^{\pr},S} \} > 1 - C_{2}L^{-d  (\frac {\rho_{1}} 2 -\rho_{2})+},
\qquad  \text{with}\\
& \Omega_{\Lambda,B,B^{\pr},S}^{\Lambda_{0}}= \bigsqcup_{i=1}^{I}\cC_{\Lambda,B\sqcup S_{i}, B^{\pr}\sqcup (S_{0}\setminus S_{i}), S\setminus S_{0}},
\end{split}\eeq
where the constants $C_{1},C_{2}$ do not depend on the scale $L$. 
 In particular, we get
\beq  \label{wegner3}
 \P \left\{ \left\{\norm{R_{\X,\Lambda}(E)} < \e^{C_{1}L^{\frac 4 3 \rho_{1}}\log L  }\right\} \cap \Omega_{\Lambda}^{(0)} \right\} > 1 - L^{-  p}.
 \eeq
\end{lemma}

\begin{proof}

Let $\cC_{\Lambda,B,B^{\pr},S}$  be  a   $(\Lambda,E)$-prepared cylinder event, consider $\Lambda_{L_{0}}\subset \Lambda$ as above, and set $B_{0}=B\cap \Lambda_{L_{0}}$,   $B_{0}^{\pr}=B^{\pr}\cap \Lambda_{L_{0}}$, and   $S_{0}= S \cap \Lambda_{L_{0}}$. Let
\beq
H_{\beps_{S_{0}}}:=  H_{B,(S,\beps_{S}),\Lambda_{L_{0}}}= H_{B_{0},(S_{0},\beps_{S_{0}}),\Lambda_{L_{0}}}=- \Delta_{\Lambda_{L_{0}}}+ V_{B_{0}} +  \sum_{s\in S_{0}} \beps_{s}(\omega) u(x-s),
\eeq
where $\beps_{S_{0}}=\{\beps_{s}\}_{s\in S_{0}}$ are  independent Bernoulli random variables, with $ \P_{\beps_{S_{0}}}$ denoting the corresponding probability measure.
All the   hypotheses of  \cite[Lemma~5.1$^{\pr}$]{BK} are satisfied by the random operator  $H(\beps_{S_{0}})$ in the box $\Lambda_{L_{0}}$.
In particular it follows from      the density condition
\eqref{densitycond2} that  $S_{0}$ is a collection of ``free sites '' satisfying  the condition in  \cite[Eq.~(5.29)]{BK} inside the box $\Lambda_{L_{0}}$.  (The fact that we have a configuration
 $ B_{0}\cup B_{0}^{\pr} \cup S_{0} \subset \J_{\Lambda}$ instead of a subconfiguration of $\Z^{d}$ is not important; only the density condition  \cite[Eq.~(5.29)]{BK}    and the fact that
$\cC_{\Lambda,B_{0},B_{0}^{\pr},S_{0}}$ is $(\Lambda_{L_{0}},E)$-prepared  matter,  the specific location of the  single-site potentials   plays no role in the analysis.)

  Thus it follows from  \cite[Lemma~5.1$^{\pr}$]{BK}  that ($L$ large)
 \beq \label{sperner}
  \P_{\beps_{S_{0}}} \left\{ \norm{R_{\beps_{S_{0}}}(E)} < \e^{C_{1} \ell_{1}^{\frac 4 3}\log \ell_{1}}\right\} > 1 - C_{2} \ell_{2}^{d}\ell_{1}^{-\frac d 2  +},
 \eeq
where the constants $C_{1},C_{2}$ do not depend on the scale $L$. In other words,
there is a subset  $Q \subset \{0,1\}^{S_{0}}$ such that
 \beq \label{sperner2}\begin{split}
& \P \{\beps_{S_{0}}\in Q\} > 1 - C_{2} \ell_{2}^{d}\ell_{1}^{-\frac d 2  +}, \quad \text{and}\\
& \norm{R_{B\cup \cX(S_{0},\eps_{S_{0}}),\Lambda_{L_{0}}}(E)} < \e^{C_{1} \ell_{1}^{\frac 4 3}\log \ell_{1}}\quad \text{for all} \quad \eps_{S_{0}}\in Q.
\end{split}\eeq

We now conclude from  \eqref{sperner2}, recalling the definitions of $\ell_{1}$ and $\ell_{2}$, that   there  exist disjoint $\Lambda$-bevents 
$\{\cC_{\Lambda,B\sqcup S_{i}, B^{\pr}\sqcup (S_{0}\setminus S_{i}), S\setminus S_{0}}\}_{i=1,2,\ldots,I}$, with each $S_{i}\subset \S_{0}$, such that we have \eqref{wegner1} and \eqref{wegner2}.
 
 Since the event $\Omega_{\Lambda}^{(1)}$ in \eqref{prepared} is a disjoint union of
 such  $(\Lambda,E)$-prepared bevents, we have, using also Lemma~\ref{lemmovepoint} as in the derivation of \eqref{qweg} (and changing $C_{1}$ slightly),  that
 \beq
 \P \left\{\left. \left\{\norm{R_{\X,\Lambda}(E)} < \e^{C_{1}L^{\frac 4 3 \rho_{1}}\log L  }\right\} \cap \Omega_{\Lambda}^{(0)} \right|\Omega_{\Lambda}^{(1)} \right\} > 1 - C_{2}L^{-d  (\frac {\rho_{1}} 2 -\rho_{2})+},
 \eeq
and hence, using the probability estimate in  \eqref{prepared}, we have
 \beq
 \P \left\{ \left\{ \norm{R_{\X,\Lambda}(E)} < \e^{ 2C_{1} L^{\frac 4 3 \rho_{1}}\log L  }\right\} \cap \Omega_{\Lambda}^{(1)} \right\} > 1 - 2 C_{2}L^{-d  (\frac {\rho_{1}} 2 -\rho_{2})+}.
 \eeq 
The desired \eqref{wegner3} follows using \eqref{rhos}.
\end{proof}

We are now ready to finish the proof of Proposition~\ref{propA}.

\begin{proof}[Proof of Proposition~\ref{propA}]
 Fix $ E \in [0,E_{0}]$.  It suffices to prove that if $L^{\prime}$ is $E$-localizing for all $L^{\prime} \in [L^{\rho_{2}},L^{\rho_{1}}]=[\ell_{2},\ell_{1}]$,  and the scale  $L$ is sufficiently large,      then  $L$ is $E$-localizing.

   Let  $\cC_{\Lambda,B,B^{\pr},S}$ be a  $(\Lambda,E)$-prepared  bevent, so there is $R^{\prime}\subset R_{0}$ with $ \#(R_{0} \setminus R^{\prime})\le K_{1} $,  such that  if $r \in R^{\prime}$ then  
$\cC_{\Lambda,B,B^{\pr},S}^{\Lambda_{\ell_{1}}(r)}$ is  a $(\Lambda,\Lambda_{\ell_{1}}(r),E)$-adapted bevent, and if $r \in R_{0} \setminus R^{\prime}$ then
$S \cap \Lambda_{\ell_{1}}(r)=\emptyset $ and $\cC_{\Lambda,B,B^{\pr}}^{\Lambda_{\ell_{1}}(r)}$ is  a $(\Lambda,\Lambda_{\ell_{1}}(r),E)$-notsobad bevent.  Recalling \eqref{nesting}, we pick $n_{0}\in \N$ such that  $\ell_{0}:= (2 n_{0} \alpha +1)\ell_{1}\sim L^{1-}$.  By geometrical considerations, we can find boxes
$\Lambda^{(j)}=\Lambda_{ (2 m_{j}n_{0} \alpha +1)\ell_{1}}(s_{j}) \subset \Lambda$,  $j=1,2,\dots,J$, 
where $J \le K_{1}$, with $m_{j}\in \{1,2, \ldots,2K_{1} \}$ and $s_{j} \in  \G_{\Lambda}^{(\ell_{1})}$ for
each  $j=1,2,\dots,J$, such that
$\mathrm{dist}(\Lambda^{(j)},\Lambda^{(j^{\pr})})\ge \ell_{0}$ if $j\ne j^{\pr}$, and for each   $r \in R_{0} \setminus R^{\prime}$ there is $j_{r} \in \{1,2,\ldots,J\}$ such that
$\Lambda_{\frac{\ell_{0}} 5}(r)\cap \Lambda \subset \Lambda^{(j_{r})}$.

Since each $\Lambda^{(j)}$ is of the form given in \eqref{nesting},  we can apply Lemma~\ref{lemwegner} to each $\Lambda^{(j)}$. Since the $\Lambda^{(j)}$ are disjoint, we can use independence of events based in different  $\Lambda^{(j)}$'s, and we may apply Lemma~\ref{lemwegner} (or its proof) to all $\Lambda^{(j)}$.
Setting $S_{0}=   \bigcup_{j=1}^{J} S\cap \Lambda^{(j)}$ and $\tilde{S}
=S \setminus S_{0}$, we conclude that there exist disjoint subsets $\{S_{q}\}_{q=1,2,\dots,Q}$ of $S_{0}$, such that for each $q=1,2,\dots,Q$  and  all $ t_{\tilde{S}} \in [0,1]^{\tilde{S}}$ we have   
\beq  \label{wegner11}
\norm{R_{B\sqcup S_{q}, (\tilde{S},t_{\tilde{S}}),\Lambda^{(j)}}(E)}< 
\e^{C_{1}L^{\frac 4 3 \rho_{1}}\log L  },\quad \text{for all} \quad  j=1,2 \ldots,J,
\eeq
and we  have the  conditional probability estimate
\beq \begin{split}\label{wegner22}
&\P\{\left. \Omega_{\Lambda,B,B^{\pr},S}^{\pr}\right|\cC_{\Lambda,B,B^{\pr},S} \} > 1 -2 K_{1} C_{2}L^{-d  (\frac {\rho_{1}} 2 -\rho_{2})+}
\qquad  \text{with}\\
&
\Omega_{\Lambda,B,B^{\pr},S}^{\pr}= \bigsqcup_{i=q}^{Q}\cC_{\Lambda,B\sqcup S_{q},B^{\pr}\sqcup (S_{0}\setminus S_{q}), \tilde{S}}.
\end{split}\eeq
By construction, each configuration in  $C_{\Lambda,B\sqcup S_{q}, \tilde{S}}$ satisfies the hypotheses of \cite[Lemma~2.14]{BK} (see also \cite[(2.22) and (2.23)]{BK}), and hence, recalling also \eqref{rhos},  we can conclude that  $C_{\Lambda,B\sqcup S_{q}, \tilde{S}}$ is a $(\Lambda,E)$-good bconfset.
Since  it is clear that $\tilde{S}$ satisfies the density condition \eqref{densitycond} in $\Lambda$, each  $\cC_{\Lambda,B\sqcup S_{q},B^{\pr}\sqcup (S_{0}\setminus S_{q}), \tilde{S}}$ is a $(\Lambda,E)$-adapted bevent.

Recalling   Lemma~\ref{lemstructure} and  the event $  \Omega_{\Lambda}^{(1)}$ in  \eqref{prepared}, we conclude the existence of  disjoint  $(\Lambda,E)$-adapted  bevents  $\{\cC_{\Lambda,B_{i},B_{i}^{\pr},S_{i}}\}_{i=1,2,\ldots,I}$, and hence of the $(\Lambda,E)$-localized event 
\beq
 \Omega_{\Lambda}= \bigsqcup_{i=1}^{I} \cC_{\Lambda,B_{i},B_{i}^{\pr},S_{i}},
\eeq
such that 
\beq \label{adaptedagain}
\P\{\left. \Omega_{\Lambda}\right|  \Omega_{\Lambda}^{(1)} \} >  1 -2 K_{1} C_{2}L^{-d  (\frac {\rho_{1}} 2 -\rho_{2})+}.
\eeq
Using the probability estimate in \eqref{prepared} and  \eqref{rhos},
we get that 
\beq
\P\{\Omega_{\Lambda} \} >  1 - L^{-p},
\eeq 
and hence the scale $L$ is $E$-localizing.

Proposition~\ref{propA} is proven.
\end{proof}

\section{The proofs of Theorems~\ref{thmpoisson} and \ref{thmpoissonbis}}
\label{sectEND}

In view of Propositions~\ref{proppoisson}, \ref{proppoissonbis}, and \ref{propA}, 
Theorems~\ref{thmpoisson} and \ref{thmpoissonbis} are a consequence of the following proposition, whose hypothesis follows from the  conclusion of  Proposition~\ref{propA}.  We recall Definition~\ref{jlocalizing}.

\begin{proposition}\label{propB} Fix  $p=\frac 3 8 d -$ and an energy $E_{0}>0$, and  suppose there is a scale $L_{0}$ and $m>0$ such that $L$ is $(E,m)$-jlocalizing for all $L \ge L_{0}$ and $E \in [0,E_{0}]$. Then   the following holds  ${\P}$-a.e.: The operator $H_\X$ has pure point spectrum  in $[0,E_0]$ with exponentially localized eigenfunctions  (exponential localization)  with rate of decay $\frac m 2$, i.e., if    $\phi$ is an eigenfunction of $H_\X$ 
with eigenvalue $E \in[0,E_0]$ we have
\begin{equation}\label{expdecay7}
 \|\chi_x \phi\| \le C_{\X,\phi} \, e^{-\frac m 2 |x|}, \quad \text{for all $x \in \R^{d}$}.
\end{equation}
Moreover, there exist   $\tau>1$ and  $s\in]0,1[$ such that for   eigenfunctions $\psi,\phi$ (possibly equal) with the same eigenvalue  $E\in[0,E_0]$ 
we have 
\begin{equation}\label{SUDEC7}
\| \chi_x\psi\| \, \|\chi_y \phi\| \le C_{\X} \|T^{-1}\psi\|\|T^{-1}\phi\| \, e^{\scal{y}^\tau} e^{-|x-y|^s}, \quad \text{for all $x,y \in \Z^{d}$}.
\end{equation}
 In particular, the eigenvalues of $H_\X$ in $[0,E_0]$ have finite multiplicity,  and  $H_\X$ exhibits dynamical localization in $[0,E_0]$, that is, for any $p>0$  we have
\begin{equation}\label{dynloc7}
\sup_t \| \scal{x}^p e^{-itH_\X} \chi_{[0,E_0]}(H_\X) \chi_0 \|^2_2 < \infty.
 \end{equation}
\end{proposition}

\begin{proof}

The fact that the hypothesis of  Proposition~\ref{propB} imply exponential localization in the interval $ [0,E_{0}]$ is proved in \cite[Section~7]{BK}.  Although their proof is written for the Bernoulli-Anderson Hamiltonian, it also applies to the Poisson Hamiltonian by proceeding as in the proof of Proposition~\ref{propA}.  When \cite[Section~7]{BK} states that a box $\Lambda$ is good at energy $E$, we should  interpret it  as the occurrence of  the $(\Lambda,E,m)$-jlocalized event $\Omega_{\Lambda}$ as in  \eqref{adapted3j}, with probability satisfying the estimate  \eqref{PEL0j}, whose existence is guaranteed by the hypothesis of Proposition~\ref{propB}.  We should rewrite such an event as in Lemma~\ref{lemlocLell} when necessary, with   $p^{\pr}=\frac 3 8 d -<p$. 
With these modifications, plus the use of Lemmas~\ref{lemqgood} and \ref{lemqgood7}  when necessary, the analysis of  \cite[Section~7]{BK} yields exponential localization for Poisson Hamiltonians.

The decay of eigenfunction correlations given in \eqref{SUDEC7}  follows for the Bernoulli-Anderson Hamiltonian  from a careful analysis of \cite[Section~7]{BK} given in \cite{GKsudec2}, and hence it also holds for the Poisson Hamiltonian  by the same considerations as above.  Finite multiplicity and dynamical localization then follow as in \cite{GKsudec2}.
\end{proof}

\begin{acknowledgement} Fran\c cois Germinet thanks the hospitality of the University of California,
Irvine, and of the University of Kentucky, where this work began.
\end{acknowledgement}


\begin{thebibliography}{AAAAA}

\bibitem[AENSS]{AENSS}  Aizenman, M., Elgart, A., Naboko, S.,   
 Schenker, J.,  Stolz, G.: Moment analysis for localization in random Schr\"odinger 
operators. Inv. Math. \textbf{163}, 343-413 (2006)
%

\bibitem[B]{B} Bourgain, J.: On localization for lattice Schr\"odinger operators involving Bernoulli variables. Geometric aspects of functional analysis. Lecture Notes in Math. \textbf{1850}, 77-99.  Berlin: Springer, 2004

\bibitem[BK]{BK} Bourgain, J., Kenig, C.: On localization in the continuous Anderson-Bernoulli model in higher dimension,  Invent. Math. \textbf{161}, 389-426 (2005)

\bibitem[CKM]{CKM}  Carmona, R.,  Klein, A.,   Martinelli, F.: {Anderson
localization for  Bernoulli and other singular potentials}.  Commun.
Math. Phys. {\bf 108}, 41-66 (1987)

\bibitem[CL]{CL}  Carmona, R,  Lacroix, J.: {\em Spectral theory of random
Schr\"odinger operators}. Boston: Bir\-kha\"user, 1990


\bibitem[CoH]{CH} Combes,  J.M.,   Hislop, P.D.: {Localization for some
continuous, random Hamiltonians in d-dimension}. J. Funct. Anal. \textbf{124},
149-180 (1994)

\bibitem[CoHK]{CHK} Combes,  J.M.,   Hislop, P.D.,  Klopp, F.: 
 H\"older continuity of the integrated density of states for some random
 operators at all energies.  IMRN \textbf{4}, 179-209 (2003)


\bibitem[CoHKN]{CHKN}  Combes,  J.M.,   Hislop, P.D., Klopp, F. Nakamura, S.:
 The Wegner estimate and the integrated density of states for some random
operators.  Spectral and inverse spectral theory (Goa, 2000),
 Proc. Indian Acad. Sci. Math. Sci. \textbf{112} , 31-53 (2002) 




\bibitem[CoHM]{CHM} Combes,  J.M.,   Hislop, P.D., Mourre, E.: Spectral averaging,
 perturbation of singular spectra, and localization.  Trans. Amer. Math. Soc.
\textbf{348}, 4883-4894 (1996)

         

\bibitem[CoHN]{CHN} Combes,  J.M.,   Hislop, P.D.,  Nakamura, S.:{ The
$\mathrm{L}^p$-theory of the spectral  shift function, the Wegner estimate and the
integrated density of states for some random operators}. 
Commun. Math. Phys. {\bf 218}, 113-130 (2001)

\bibitem[DV]{DV} Donsker, M., Varadhan, S.R.S.: Asymptotics for the Wiener sausage.
Comm. Pure Appl. Math. \textbf{28}, 525-565 (1975)

\bibitem[DrK]{vDK} von Dreifus, H.,  Klein, A.:  {A new proof of localization in
the Anderson tight binding model}.  Commun. Math. Phys. \textbf{124},
285-299  (1989) 

\bibitem[FK]{FK} Figotin, A.,  Klein, A.: {Localization of classical waves I:
Acoustic waves}.  Commun. Math. Phys. {\bf 180}, 439-482 (1996)

\bibitem[FiLM]{FLM} Fischer, W., Leschke, H., M\"uller, P.:  Spectral localization by Gaussian random potentials in multi-dimensional continuous space.
J. Stat. Phys. \textbf{101}, 935-985 (2000) 

\bibitem[FrMSS]{FMSS} Fr\"ohlich, J.:    Martinelli, F.,  Scoppola, E., 
Spencer, T.:  {Constructive proof of localization in the Anderson tight
binding model}. Commun. Math. Phys. {\bf 101}, 21-46 (1985)


\bibitem[FrS]{FS}  Fr\"ohlich, J.,  Spencer, T.: {Absence of diffusion with
Anderson tight binding model for large disorder or low energy}. Commun.
Math. Phys. {\bf 88}, 151-184 (1983)

\bibitem[GHK1]{GHK}  Germinet, F., Hislop, P., Klein, A.: Localization for the Schr\"odinger operator with a Poisson random potential. C.R. Acad. Sci. Paris Ser. I \textbf{341}, 525-528 (2005) 

\bibitem[GHK2]{GHK2}  Germinet, F., Hislop, P., Klein, A.: 
Localization at low energies for attractive Poisson random Schr\"odinger operators.
CRM Proceedings \& Lecture Notes. To appear

\bibitem[GK1]{GK1}  Germinet, F., Klein, A.: {Bootstrap Multiscale Analysis
and Localization in Random Media}. Commun. Math. Phys. \textbf{222}, 415-448 (2001)


\bibitem[GK2]{GK2}  Germinet, F,  Klein, A.: Operator kernel estimates for  functions of  generalized Schr\"odinger operators.
Proc. Amer. Math. Soc. \textbf{131},  911-920  (2003)


\bibitem[GK3]{GKgafa} Germinet, F.,  Klein, A.: {Explicit finite volume criteria for
localization in random media and applications}.  Geom. Funct. Anal. \textbf{13}, 
1201-1238 (2003)


\bibitem[GK4]{GKsudec} Germinet, F., Klein, A.: New characterizations of the region of complete localization for random Schr\"odinger operators.  J. Stat. Phys. \textbf{122}, 73-94 (2006)


\bibitem[GK5]{GKsudec2} Germinet, F., Klein, A.: In preparation

\bibitem[HM]{HM}   Holden, H., Martinelli, F.:  On absence of diffusion
near the bottom of the spectrum for a random Schr\"odinger operator.
 Commun. Math. Phys. {\bf 93}, 197-217 (1984)
 
\bibitem[K]{King} Kingman, J.F.C.:
    \emph{Poisson processes}. New York: The Clarendon Press Oxford University Press, 1993
    
   
\bibitem[Ki]{Kir} Kirsch, W.: Wegner estimates 
and Anderson localization for alloy-type potentials.
 Math. Z. \textbf{221},  507--512 (1996) 


\bibitem[KiM]{KM}  Kirsch, W.,   Martinelli, F.: {On the ergodic properties of 
the spectrum of general random operators}.  J. Reine Angew. Math. {\bf
334},  141-156  (1982)


\bibitem[KiSS]{KSS} Kirsch, W.,   Stollmann, P., Stolz, G.: {Localization
for random perturbations of periodic Schr\"odinger
 operators}.  Random Oper. Stochastic Equations {\bf 6}, 241-268
(1998) 

\bibitem[Kl]{Kle} Klein, A.:   Multiscale analysis and localization of random operators.
 In \emph{Random Schr\"odinger operators: methods, 
 results, and perspectives}.  Panorama \& Synth\`{e}se, Soci\'{e}t\'{e}
 Math\'{e}matique de France.  To appear



\bibitem[Klo1]{Klop93}  Klopp,  F.: {Localization for semiclassical continuous random Schr\"odinger operators. II. The random displacement model}.
{Helv. Phys. Acta} \textbf{66}, 810-841  (1993)

\bibitem[Klo2]{Klop95} Klopp,  F.: {Localization for continuous random
Schr\"odinger operators}. Commun. Math. Phys. {\bf 167}, 553-569
(1995)

\bibitem[Klo3]{Klop97} Klopp F.:  A low concentration asymptotic expansion for the density of  states of a random Schr\"odinger operator with Poisson disorder.  J. Funct. Anal. {\bf 145} 267--295  (1997)

\bibitem[Klo4]{Klop02}  Klopp F.: Weak disorder localization and Lifshitz
tails: continuous Hamiltonians. Ann. I.H.P. {\bf 3}, 711-737 (2002)

\bibitem[KloP]{KP} Klopp, F.,  Pastur, L.:
   {Lifshitz tails for random Schr\"odinger operators with negative singular Poisson potential}.
{Comm. Math. Phys.} \textbf{206}, 57-103 (1999)
    
\bibitem[LMW]{LMW}  Leschke, H.,  M{\"u}ller, P., Warzel, S.:
 A survey of rigorous results on random Schr\"odinger operators for amorphous solids.
Markov Process. Related Fields   \textbf{9}, 729-760 (2003)

 
 \bibitem[LiGP]{LGP} Lifshits, I.M., Gredeskul, A.G., Pastur, L.A.: 
\emph{Introduction to the Theory of Disordered Systems}.
New York:  Wiley-Interscience, 1988
 
 
 
\bibitem[MS]{MS}  Martinelli, F.,  Scoppola, E.: {Introduction to the
mathematical theory of  Anderson localization}. Riv. Nuovo Cimento {\bf 10},
1-90 (1987)

 
 \bibitem[PF]{PF}  Pastur, L.,  Figotin, A.: {\em Spectra of Random and
Almost-Periodic Operators}.  Heidelberg: Springer-Verlag, 1992




\bibitem[R]{Reiss} Reiss, R.-D.: A course on point processes. New York:
Springer-Verlag, 1993

\bibitem[Ro]{Ro} Robbins, H.:  A remark on Stirling's formula. Amer. Math. Monthly {\bf 62} , 26--29 (1955)

\bibitem[S]{Si}  Simon, B.: {Schr\"odinger semi-groups}.  Bull. Amer.
Math. Soc. {\bf 7},  447-526 (1982)

\bibitem[SW]{SW}   Simon, B., Wolff, T.:  Singular continuum spectrum under
rank one perturbations and localization for random Hamiltonians.
Commun. Pure. Appl. Math. {\bf 39}, 75-90 (1986)




\bibitem[Sp]{S}  Spencer,  T.: {Localization for random and
quasiperiodic potentials}.    J. Stat. Phys. {\bf 51}, 1009-1019 (1988)

\bibitem[St1]{St99} {Stollmann, P.}:
   {Lifshitz asymptotics via linear coupling of disorder}.
{Math. Phys. Anal. Geom.} \textbf{2}, 279-289 (1999)

\bibitem[St2]{St2} {Stollmann, P.}:
Wegner estimates and localization for continuum Anderson models with some
singular distributions.
Arch. Math. (Basel) \textbf{75}, 307-311 (2000)

    
\bibitem[Sto]{Stolz} Stolz, G.: Localization for random Schr\"odinger operators with Poisson potential. Ann. Inst. H. Poincar\'e Phys. Th\'eor. 
\textbf{63} , 297-314 (1995)

\bibitem[Sz]{Sz} {Sznitman, A.-S.}:
 \emph{Brownian motion, obstacles and random media}. Berlin:
{Springer-Verlag}, {1998}

\bibitem[U]{U} Ueki, N.:  Wegner estimates and localization for
 Gaussian random potentials. Publ. Res. Inst. Math. Sci.
 \textbf{40}, 29-90 (2004)


\end{thebibliography}
\end{document}